\documentclass[aps,prd,preprint,groupedaddress,nofootinbib,showpacs]{revtex4}

\usepackage{bm}
\usepackage{slashed}
\usepackage{amstext,amsmath,amsfonts,amssymb}
\usepackage[latin1]{inputenc} 
\usepackage[dvips]{graphicx}

\newcommand{\gev}{\text{GeV}}
\newcommand{\tev}{\text{TeV}}
\newcommand{\fb}{\text{fb}}

\newcommand{\U}{{\cal U}}

\newcommand{\rar}{\rightarrow}

\def\d{\:\text{d}}
\def\bar{\overline}
\def\lang{\left\langle}
\def\rang{\right\rangle}
\def\lvert{\left\vert}
\def\rvert{\right\vert}

\newcommand\OO{\mathcal O}
\newcommand{\qu}[1]{`#1'}
\newcommand{\eref}[1]{Eq.~(\ref{#1})}

\newcommand{\bcen}{\begin{center}}
\newcommand{\ecen}{\end{center}}
\newcommand{\be}{\begin{equation*}}
\newcommand{\ee}{\end{equation*}}

\def\openone{\leavevmode\hbox{\small1\normalsize\kern-.33em1}}

\begin{document}

\title{Unparticle Self-Interactions at the Large Hadron Collider}

\author{Johannes Bergstr\"om}
\email{johbergs@kth.se}

\author{Tommy Ohlsson}
\email{tommy@theophys.kth.se}

\affiliation{Department of Theoretical Physics, School of
Engineering Sciences, Royal Institute of Technology (KTH) --
AlbaNova University Center, Roslagstullsbacken 21, 106 91 Stockholm,
Sweden}

\date{\today}

\begin{abstract}
We investigate the effect of unparticle self-interactions at the Large
Hadron Collider (LHC). Especially, we discuss the three-point
correlation function, which is determined by conformal symmetry up to
a constant, and study its relation to processes with four-particle
final states. These processes could be used as a favorable way to look for unparticle physics,
and for weak enough couplings to the Standard Model, even the only way.
We find updated upper bounds on the cross sections for
unparticle-mediated $4\gamma$ final states at the LHC and novel upper
bounds for the corresponding $2\gamma 2\ell$ and $4\ell$ final
states. The size of the allowed cross sections obtained are comparably
large for large values of the scaling dimension of the unparticle
sector, but they decrease with decreasing values of this parameter. In
addition, we present relevant distributions for the different final
states, enabling the possible identification of the unparticle scaling
dimension if there was to be a large number of events of such final
states at the LHC.
\end{abstract}


\maketitle


\section{Introduction}
\label{sec:intro}

The Standard Model (SM) of particle physics has to date been extremely
successful yielding consistent results and most accurate predicting
properties of particles and their interactions. However, in its
minimal version, it does not contain massive neutrinos, gravity,
and/or dark matter. Although it should be fairly easy to include
massive neutrinos as well as neutrino oscillations in a minimally
extended version of the SM, one of the most important particles of the
SM has not yet been experimentally confirmed, namely the Higgs boson,
which is responsible for giving masses to the particles of the
SM. Therefore, the Large Hadron Collider (LHC) has been constructed
and will hopefully be able to discover it. In the connection with the
process of constructing the LHC, a lot of alternative pictures have
been proposed, including for example unparticle physics that was
introduced by Howard Georgi in 2007 \cite{Georgi:2007ek}. Thus, in the
last couple of years, several studies of the phenomenology of
unparticle physics have appeared in the literature. In this work, we
will focus on studying the effects of unparticle self-interactions at
colliders, and especially the LHC. In general, our investigation will
be based on the unparticle three-point correlation function, and in
particular, we will discuss the phenomenology of unparticle
self-interactions in processes with different four-particle final
states such as four photons ($4\gamma$), two photons and two charged
leptons ($2\gamma 2\ell$), and four charged leptons ($4\ell$) as well
as final states including neutrinos, i.e., $2\gamma \nu \bar\nu$ and
$2\ell \nu \bar\nu$.

Previously, an investigation of unparticle self-interactions that is of great
importance for this work has been presented in the following paper: In
Ref.~\cite{Feng:2008ae}, collider signals of unparticle self-interactions 
especially with four-photon final states are investigated, which lead to
upper bounds on cross sections using Tevatron data that follow from
restricting the unparticle self-interactions. In addition, Strassler
\cite{Strassler:2008bv} has discussed self-interactions in general and
Georgi and Katz \cite{Georgi:2009xq} have discussed the production of
unparticles through the amputated three-point function.

Our work is organized as follows. In Sec.~\ref{sec:model}, we
introduce our model based on a conformal hidden sector coupled to the
SM and also present the specific couplings to the SM. Then, in
Sec.~\ref{sec:tpf}, we explicitly derive the unparticle three-point
correlation function and discuss the general form of the four-point
correlation function. Next, in Sec.~\ref{sec:phen}, we present the
phenomenology of our model concerning different four-particle final
states mediated by the unparticle three-point function at the LHC. By
using existing Tevatron data, we constrain the constant appearing in
the three-point function, enabling determination of the maximum
allowed cross sections at the LHC. Furthermore, relevant distributions
are given, which could be used to identify unparticle physics as a
possible interpretation of future data in the event that a large
number of events with four-body final states were to be measured at
the LHC. Finally, in Sec.~\ref{sec:summary}, we give a short summary
of our work as well as our main conclusions.

\section{The Model}
\label{sec:model}

The most famous examples of four-dimensional conformal field theories,
which could serve as hidden sectors, are theories of so-called
Banks--Zaks type \cite{Banks:1981nn}. These are gauge theories with
the number of fermions chosen such that an infrared stable fixed point
emerges at two loops. In addition, one can have supersymmetric
versions of these types of theories and even calculate the fraction of
them being conformal in the infrared regime \cite{Ryttov:2007sr}.

In general, unitary four-dimensional scale-invariant quantum field
theories are also conformally invariant with no known counterexamples
\cite{Polchinski:1987dy,Nakayama:2007qu,Grinstein:2008qk}. Unitarity demands that the
scaling dimension of an operator with Lorentz spins $j_1$ and $j_2$ is
$ d \geq j_1 + j_2 + 2 - \delta_{j_1 j_2,0}$ \cite{Mack:1975je}. This
translates into $ d_{S} \geq 1$, $d_{Sp} \geq 3/2$, and $d_{V} \geq 3$
for Lorentz scalar, spinor, and gauge invariant vector operators,
respectively. However, much of the phenomenology of unparticle
physics dealing with vector operators are performed for values
violating these bounds. One also finds that the fields satisfy
equations of motion for free fields and conservation laws if and only
if the lower bounds are saturated. For vector operators, this includes
the existence of zero-norm descendants $ \partial_{\mu} V^{\mu}(x) = 0
$, which does not hold away from the unitarity bound.

The basic scheme for our model is the following. There is a hidden
sector of Banks--Zaks type that couples to the SM through a
\qu{messenger field} with a large mass $M$, which simply means that
both the SM and the hidden sector couple to this field. Below the
scale $M$, one can use effective field theory and integrate out the
heavy field, whereby one ends up with effective operators suppressed
by powers of $M$ on the form\footnote{Our notation will follow that of
  Ref.~\cite{Bander:2007nd}.}
\begin{equation}
\label{eq:effcoupling} c_n^i \frac{1}{M^{n+d_{UV} - 4}}\OO_{n}^i \OO_{UV},
\end{equation}
where $c_n^i$ are dimensionless constants and $\OO_{n}^i$ and
$\OO_{UV}$ are local operators constructed out of the SM and hidden
sector fields with mass dimensions $n$ and $d_{UV}$, respectively.
The hidden sector becomes conformal at the scale
$\Lambda_{\U}$,\footnote{This scale can for example be defined as the
  scale, where the coupling has reached $\sqrt{2/3}$ of its
  fixed-point value.} for example, by quantum effects via dimensional
transmutation. The operator $\OO_{UV}$ changes into
$\Lambda_{\U}^{d_{UV}-d} \OO_\U$, where $\OO_{\U}$ has scaling and
mass dimension $d$. This implies that the couplings in
\eref{eq:effcoupling} have to be replaced by
\begin{equation}
c_n^i \frac{\Lambda_{\U}^{d_{UV}-d}}{M^{n+d_{UV} - 4}}\OO_{n}^i
\OO_{\U} = c_n^i \frac{1}{\Lambda_n^{n+d - 4}} \OO_{n}^i \OO_{\U},
\end{equation}
where the effective scale $\Lambda_n$ is given by
\begin{equation}
\Lambda_n^{n+d - 4} = \frac{M^{n+d_{UV} - 4}}{\Lambda_{\U}^{d_{UV}-d}}.
\end{equation}
Note that $M$, $\Lambda_\U$, $d$, and $d_{UV}$ are parameters of the
messenger and hidden sector, while $n$ will depend on what SM operator
is under consideration. Thus, $\Lambda_n$ will depend on $n$ (and
hence the subscript), i.e., operators of different mass dimension will
couple with different strengths. In fact, these different energy
scales counteract the effective field theory intuition that
higher-dimensional operators are more highly suppressed
\cite{Bander:2007nd}. In addition, note that the elimination of the
heavy field induces contact interactions between generic SM fields
such as
$$
\OO_{n}^i\OO_{n'}^{i'}, \; \OO_{n}^i\partial^2 \OO_{n'}^{i'}, \; \cdots,
$$
suppressed by powers of $M$. These interactions have the potential
to drown any unparticle effects \cite{Grinstein:2008qk}, at least for
values of $d$ much larger than 1, and thus could give the most
stringent constraints on the parameter space relevant for unparticle
physics.

Since the scaling dimension of the scalar unparticle operator exceeds
one and all the SM operators except one that it can couple to all have
mass dimension $n \geq 3$, these couplings are all irrelevant. The
only coupling that can possibly be relevant is the Higgs-unparticle
coupling
\begin{equation}
\frac{\Lambda_\U^{d_{UV}-d}}{M^{d_{UV}-2}} H^{\dagger} H \OO_{\U} =
\frac{1}{\Lambda_2^{d-2}} H^{\dagger} H \OO_{\U}.
\end{equation}
For $d<2$, this coupling is indeed relevant and can significantly
alter the low-energy physics of the unparticle sector. Once the Higgs
field develops a vacuum expectation value $v$ (giving
$H^{\dagger} H \rar v^2$), this operator introduces a scale into the
conformal field theory and will cause the theory to become
nonconformal at the energy scale $\Lambda_{\slashed{\U}}$
\cite{Fox:2007sy}, where
\begin{equation}
\Lambda_{\slashed{\U}}^{4-d} = \Lambda_2^{2-d} v^2.
\end{equation}
Below the scale $\Lambda_{\slashed{\U}}$, the hidden sector presumably
becomes a conventional particle sector. For consistency, one requires
$\Lambda_{\slashed{\U}} \leq \Lambda_{\U}$ and the two scales to be
reasonably well separated in order to give a window, where the sector
is conformal.

The standard way to treat the breaking of conformal symmetry is to
introduce a mass gap in the spectral density. In the absence of fine
tuning, this mass gap is at least several GeVs. This implies that
there are no long-range forces from unparticle exchange and that
precision constraints in practice disappear. Thus, low-energy
experiments are not sensitive to unparticle effects, and these are
most easily probed at colliders. If the energy of the process in
question is high enough, one can simply ignore the mass gap in the expression for the propagator. In
Sec.~\ref{sec:phen}, the unparticle three-point function will be used
to study possible signals at the LHC. Although it is unclear how to
implement the breaking of conformal invariance into this function, it
is a reasonable assumption that one can ignore the existence of the
mass gap when calculating the three-point function as long as \emph{all} the momentum invariants [$q_1^2$,
$q_2^2$, and $(q_1+q_2)^2$] in the function are much larger than the
conformal symmetry breaking scale. Since this will generally
be the case at the LHC, this subject will not be discussed any
further.

Covariance under conformal transformations determines the (exact)
two-point functions of primary scalar operators $\OO_i$ with scaling
dimensions $d_i$ to be
\begin{equation}
\label{eq:two-point}
\lang \OO_i(x) \OO_j(0) \rang \propto \frac{\delta_{d_i d_j}}{(-x^2 +
{\rm i} \epsilon)^{d_i}} \propto \delta_{d_i d_j} \int \frac{\d
p^4}{(2 \pi)^4} (-p^2-{\rm i} \epsilon)^{d_i-2}e^{-{\rm i}p \cdot x},
\end{equation}
which give the expressions for the momentum space propagators that can
enter in processes with internal unparticle lines. The form of the
propagator can give unusual kinematical distributions due to the
unusual dependence on $p^2$ and the unusual phase, i.e., $e^{-{\rm i}
\pi d}$, which can lead to strange interference effects with photon
and $Z$ boson propagators
\cite{Cheung:2007ap,Cheung:2007zza,Georgi:2007si}. Furthermore, the
phase space is given by the imaginary part of the propagator
\begin{equation}
\label{eq:phasesapce}
\d \Phi \propto \theta (P^0) \theta(P^2) (P^2)^{d-2} \d^4 P .
\end{equation}
The real proportionality constant (generally denoted by $A_d$)
corresponds to field normalization and is most commonly chosen such
that the phase space interpolates that of $d$ massless particles
\cite{Georgi:2007ek}. The unparticle phase space will enter in
processes with an unparticle line in the final state. However, for
$d>2$, it is ultraviolet sensitive and one sometimes finds singular
behavior. For example, the energy density at finite temperature and
some cross sections are proportional to $2-d$
\cite{Cacciapaglia:2007jq,Rajaraman:2008qt}. This together with the
lower bound on the scaling dimension from unitarity implies that one
usually restricts the attention to the range $1\leq d \leq 2$ for
scalars.

Finally, it is generally assumed that an unparticle does not decay
into SM particles. Thus, the unparticle will simply leave any detector
undetected and lead to missing energy signals. However, the couplings
to SM fields can through loops give a width to the unparticle
propagator \cite{Rajaraman:2008bc,Delgado:2008gj}, which means that
the unparticle can, depending on model parameters, decay back into SM
particles. For an overview of the phenomenology of unparticle physics,
see for example Ref.~\cite{Cheung:2008xu} and references therein.

Higher-order processes in the unparticle sector will enter through the
unparticle $n$-point correlation function for $n\geq 3$. Some general
ideas can be found in Ref.~\cite{Strassler:2008bv}, and in
Ref.~\cite{Georgi:2009xq}, the production of unparticle stuff through
its self-interactions has been studied. In this paper, we will follow
Ref.~\cite{Feng:2008ae} and look at the implications for LHC physics
of the unparticle three-point function mediating interactions between
SM particles.

\subsection{Couplings to the Standard Model}
\label{sec:SMcouplings}

In order to calculate observables, one needs explicit
expressions for the couplings of the unparticle operator
to the SM. Initially, the only constraints on the SM operators
are that the interaction terms form Lorentz scalars.
To reduce the number of possible interactions, one
can assume that the unparticle operator is gauge invariant, and thus,
the SM operators have to be so as well. All couplings of scalar, spinor,
and vector unparticle operators to gauge invariant SM operators with
dimensions less than or equal to four have been given in
Ref.~\cite{Chen:2007qr}.

For a scalar unparticle field with scaling dimension $d$, the leading
order (in $\Lambda$) interaction with SM gauge bosons is through
\cite{Bander:2007nd}
\begin{equation}
\label{eq:scalar-gauge}
\frac{c_4^{F}}{\Lambda_4^d} F_{\mu \nu}^a F^{a \mu \nu} \OO_\U
\end{equation}
and with fermions after electroweak symmetry breaking through
\begin{equation}
\label{eq:scalar-fermion}
\frac{e c_4^f}{\Lambda_4^d} v \bar{f}_L f_R \OO_\U + \frac{e
c^{f_{L,R}}_4}{\Lambda_4^d} \bar{f}_{L,R} \gamma_\mu f_{L,R}
\partial^\mu \OO_\U,
\end{equation}
where the electromagnetic coupling has been pulled out for
convenience. The first of these two operators is proportional to
$v$. Integrating the second one by parts, the vector contribution
vanishes, while the axial-vector contribution is proportional to the
fermion mass $m_f$. Since $m_f \ll v$ for all fermions except the top
quark, the first interaction term will dominate for $c_4^f =
O(c^{f_{L,R}}_4)$. Thus, the Feynman rules for the interactions of photons, gluons, and
fermions with a scalar unparticle field are given by
\begin{align}
\OO_{\U} gg, \OO_{\U} \gamma \gamma \:\: \text{vertex:} & \;\;\;\;\;\;\;\;  {\rm i} \frac{ 4 c_{g,\gamma}}{\Lambda_4^d}(-p_a \cdot p_b g^{\mu \nu} + p^\nu_a p^\mu_b ), \\
\OO_{\U} \bar{f}f \:\: \text{vertex:} & \;\;\;\;\;\;\;\;  {\rm i} \frac{e c_4^f}{\Lambda_4^d} v P_{R}, 
\end{align} 
respectively, which will be used to study unparticle phenomenology in
the rest of this work.

For primary vectors, the interaction with SM fermions would be through
\begin{equation}
\frac{c^{f_{L,R}}_3}{\Lambda_3^{d-1}} \bar{f}_{L,R} \gamma_\mu f_{L,R} \OO^\mu_\U,
\end{equation}
which would not be suppressed as the corresponding operator in
\eref{eq:scalar-fermion}. A vector unparticle field cannot couple
directly to gauge bosons, since the number of uncontracted Lorentz
indices will always be odd. However, it can couple to them through a
term like $H^\dagger D_\mu H \OO_{\U}^\mu$. A gauge invariant unparticle spinor can only couple in the same way as
a right-handed neutrino, and thus, it is possible to replace the
ordinary right-handed neutrino with a conformal one \cite{vonGersdorff:2009is}.

\section{The Unparticle Three-Point Correlation Function}
\label{sec:tpf}

In this section, we will investigate the scalar unparticle three-point
correlation function. Then, in Sec.~\ref{sec:phen}, we will make use
of it to study unparticle self-interactions at the LHC. In general,
the three-point function can involve fields with different spins
\cite{Osborn:1993cr}, but here we will only consider scalars. The
three-point functions of scalar fields $\OO_i$ with scaling dimensions $d_i$ are constrained by conformal
invariance as
\begin{equation}
\label{eq:position TPF}
\lang \OO_1(x_1) \OO_2(x_2) \OO_3(x_3) \rang =
C_{123} \frac{1}{(x^2_{12})^{(d_1 + d_2 - d_3)/2} (x^2_{23})^{(d_2 + d_3
- d_1)/2} (x^2_{13})^{(d_1 + d_3 - d_2)/2}},
\end{equation}
for some constant $C_{123}$ and where $x_{ij} = x_i - x_j$.

\begin{figure}[ht!]
\centering
\includegraphics[width=.512\textwidth]{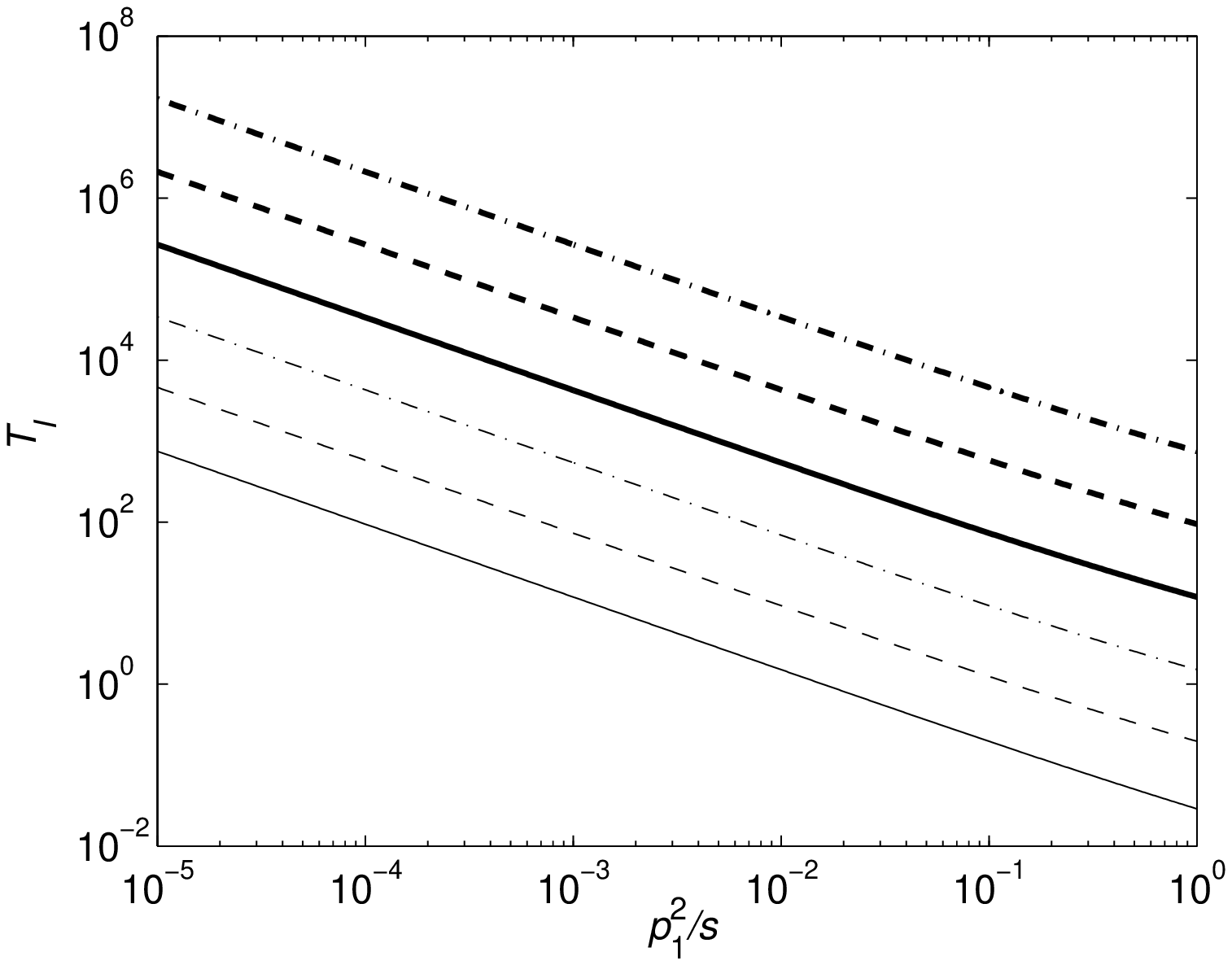}
\includegraphics[width=.512\textwidth]{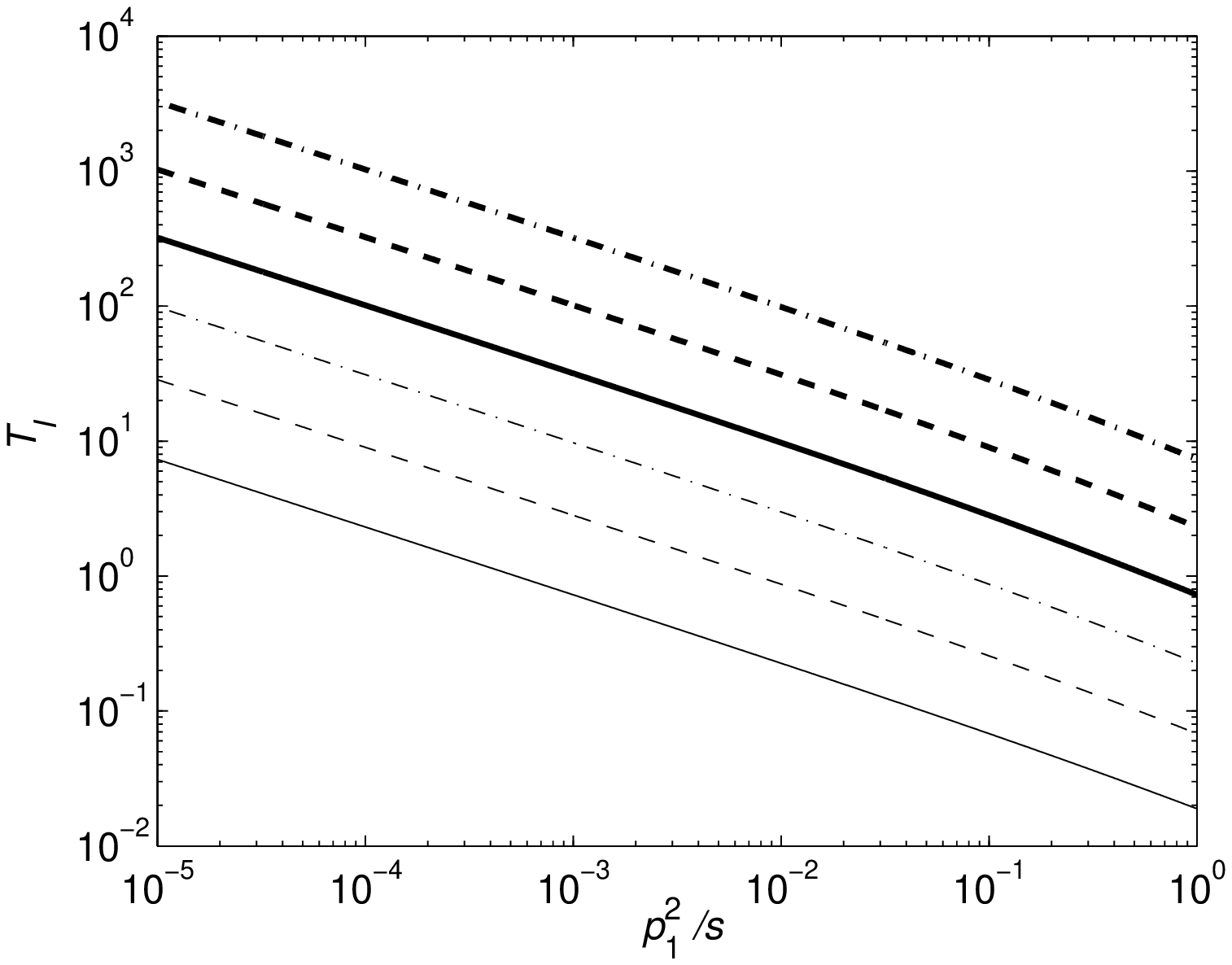}
\includegraphics[width=.512\textwidth]{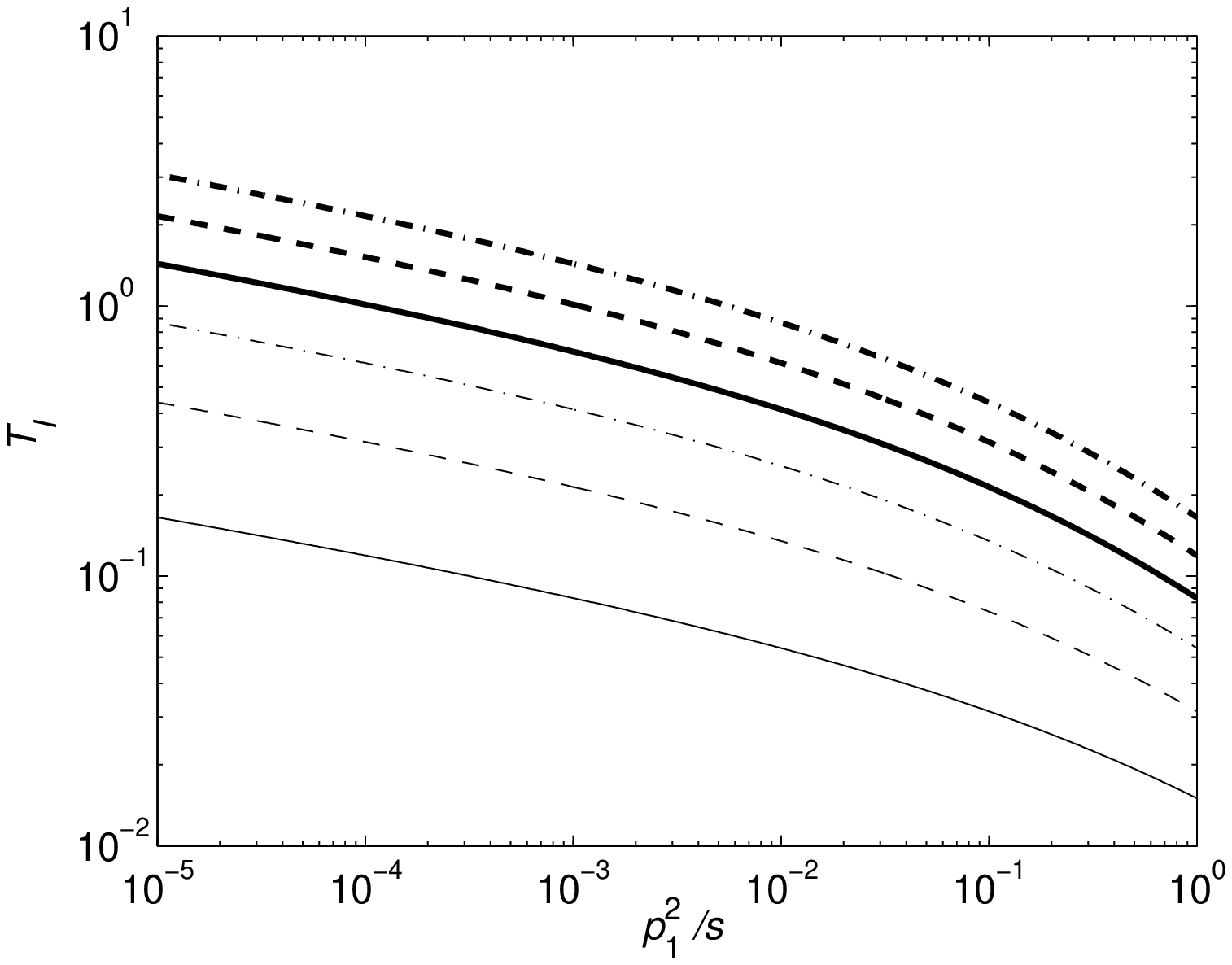}
\caption{The function
  $T_I(p_1^2/s,p_2^2/s)$ for $d=1.1$ (top panel), $d=1.5$ (middle panel), and $d=1.9$ (bottom panel). The
  graphs are shown as a function of $p_1^2/s$ for the following six discrete values: $p_2^2/s = 10^{-5}$ (thick, dash-dotted), $10^{-4}$ (thick, dashed), $10^{-3}$ (thick, solid), $10^{-2}$ (thin, dash-dotted), $10^{-1}$ (thin, dashed), and 1 (thin, solid).}\label{fig:TPF}
\end{figure}

To calculate amplitudes, one needs the Fourier transform of the
three-point function. This can be performed in a general number of
dimensions $D$ and for different values of the scaling dimensions
$d_i$ of the three fields, but we will be interested in the case
$D=4$ only and $\OO_1 = \OO_2 = \OO_{\U}$ and $\OO_3 = \OO_{\U}^\dagger$, yielding $
d_1 = d_2 =d_3 = d$. Then, Eq.~(\ref{eq:position TPF}) gives
\begin{equation}
\label{eq:momentum TPF}
\Gamma_3(p_1,p_2;d) = C_{d} \int \d^4 x_1 \d^4 x_2
\frac{1}{(x_{12}^2)^{d/2} (x_1^2)^{d/2} (x_2^2)^{d/2}} e^{{\rm i} p_1
\cdot x_1} e^{{\rm i} p_2 \cdot x_2},
\end{equation} 
for the Fourier transform of the three-point function, where $C_d$ is
a constant. Insertion of the resolution of the identity in the form
\begin{equation} 1 = \int \d^4 z
\delta^{(4)}[z - (x_1- x_2)] = \int \frac{\d^4 q}{(2 \pi)^4} \d^4 z
e^{-{\rm i} q\cdot [z-(x_1-x_2)]}, \end{equation}
yields
\begin{equation}
\label{eq:TPFmomint}
\Gamma_3(p_1, p_2; d) = C_{d} \int \frac{\d^4 q}{(2 \pi)^4}
\Gamma_2(q;d/2) \Gamma_2(p_1-q;d/2) \Gamma_2(p_2 + q;d/2),
\end{equation}
up to an overall $d$-dependent constant, which is absorbed into
$C_d$. Here \mbox{$\Gamma_2(k;g) \propto (-k^2-{\rm i}
\epsilon)^{g-2}$} is the propagator of the scalar field with scaling
dimension $g$ as in Eq.~(\ref{eq:two-point}). Since this resembles a
standard loop integral, one can introduce Feynman parameters and end
up with the result
\begin{equation}
\label{eq:3point}
\Gamma_3(p_1, p_2;d) = -{\rm i} e^{-{\rm i} \frac{3d}{2} \pi } C_d
\left(\frac{1}{s}\right)^{4-3d/2}T_I(p_1^2/s,p_2^2/s),
\end{equation} 
where
\begin{align}
T_I(p_1^2/s,p_2^2/s) =
\frac{\Gamma(4-3d/2)}{\Gamma(2-d/2)^3} \frac{1}{(4 \pi)^{2}} \int \d
y_1 \d y_2 \d y_3 \delta(y_1+y_2+y_3-1) & \notag \\ \times
\left(\frac{1}{\Delta}\right)^{4-3d/2} \left(y_1y_2y_3\right)^{1-d/2},
\end{align}
$s=(p_1+p_2)^2$ and $\Delta = x_1 x_2 p_2^2/s + x_1 x_3 p_1^2/s + x_2
x_3$. Now, integration with respect to one of the Feynman parameters,
say $y_3$, can be performed trivially due to the $\delta$ function,
leaving the integrations
\begin{equation}
\int\limits_0^1 \d y_1 \notag
\int\limits_0^{1-y_1} \d y_2 \cdots.
\end{equation}
Next, we perform the following change of variables
\begin{equation}
y_1 = 1-\omega , \;\;\;\;\; y_2 = \omega \rho
\end{equation}
with the Jacobian being equal to $\omega$. The resulting integral is
over the unit square and the integration with respect to $\omega$ can
be performed analytically, yielding the final result
\begin{align}
T_I\left(A,B\right) &= \frac{\Gamma(4-3d/2)}{\Gamma(2-d/2)^3} \frac{1-d/2}{16 \pi \sin(d \pi/2)} \int\limits_0^1 \d \rho [(1-\rho) \rho]^{1-\frac{d}{2}} \left(\frac{1}{A \rho +B -B \rho}\right)^{4-\frac{3 d}{2}} \nonumber\\& \times {}_2F_1 \left[\frac{d}{2},4-\frac{3 d}{2},2,\frac{A \rho +B -B \rho + \rho (\rho -1)}{A \rho +B -B \rho}\right]
\end{align}
with ${}_2F_1$ being the hypergeometric function. The integrand is
singular at both endpoints, but with an appropriate integration
method, it can be integrated numerically. The result for three
different values of $d$, i.e., 1.1, 1.5, and 1.9, are shown in
Fig.~\ref{fig:TPF}, which were also calculated in
Ref.~\cite{Feng:2008ae} for $d = 1.1$ and $d = 1.9$. Just as in the
case of the propagator, this is the exact three-point function (before
coupling to the SM).

If one requires scale invariance only and not full conformal invariance of the hidden sector, the three-point function is not unique. Using self-couplings of continuous mass fields, it is possible to obtain nonunique three-point functions with simple expressions in momentum space \cite{Deshpande:2008ra}. However, because of the nonuniqueness of these three-point functions, this approach is less suitable for phenomenological studies.

Similarly, the four-point function has the general form
\cite{Ginsparg:1988ui,DiFrancesco:1997nk}
\begin{equation}
\lang \OO_1(x_1) \OO_2(x_2) \OO_3(x_3) \OO_4(x_4) \rang =
\rho\left(\frac{x_{12}^2 x_{34}^2}{x_{13}^2 x_{24}^2},\frac{x_{12}^2
x_{34}^2}{x_{14}^2 x_{23}^2}\right) \prod_{i<j}(x_{ij}^2)^{\sum d_i/6-(d_i+d_j)/2},
\end{equation}
for some function $\rho$ that cannot be determined by conformal
covariance alone. For a single field of scaling dimension $d$, one
finds
\begin{equation}
\lang \OO(x_1) \OO(x_2) \OO(x_3) \OO^\dagger(x_4) \rang =
\frac{\rho'\left(\frac{x_{12}^2 x_{34}^2}{x_{13}^2
x_{24}^2},\frac{x_{12}^2 x_{34}^2}{x_{14}^2 x_{23}^2}\right)
}{(x_{13}^2)^d (x_{24}^2)^d},
\end{equation}
for $\rho'$ related to $\rho$, which, in principle, can be used to
study six-body final states. For this, one would have to assume a
specific form of $\rho'$ and then perform a Fourier transform of the
result. This analysis, however, is beyond the scope of the present work.

\section{Phenomenology}
\label{sec:phen}

In this section, the main results of this work will be presented.
These results will be based on an unparticle-induced effect
qualitatively different from those of unparticle production and
effects mediated through the propagator. This is the effect of
unparticle \emph{self-interactions}, first introduced by Feng,
Rajaraman, and Tu \cite{Feng:2008ae}, which for example can mediate
processes with four-body final states at the LHC. While the authors of
that paper considered the $4 \gamma$ final state only and the
experimental bound on the unparticle self-interaction strength using
Tevatron data in the same channel, we will in this work consider also
the $2 \gamma 2\ell$ and $4\ell$ final states at the LHC, including
those final states in which two of the leptons are neutrinos. In
addition, the $4\ell$ channel at the Tevatron will be used to give
more stringent bounds on the unparticle self-interaction strength and
thus also on the LHC cross sections.

\subsection{Unparticle Self-Interactions at the LHC}

As discussed in Sec.~\ref{sec:tpf}, three- and higher-point functions
are in general nonvanishing in conformal field theories. These
higher-point functions can mediate processes such as
$$
\Omega\bar{\Omega} \rar \U \rar \U \cdots \U,
$$
where $\Omega$ is any SM particle, $\bar{\Omega}$ is its antiparticle,
and the number of unparticles are two or more. If the conformal sector
is strongly coupled, the creation of additional high-energy
unparticles does not suppress the rate. This is in marked contrast to
all SM processes where processes with additional high-energy particles
are strongly suppressed.

\begin{figure}[ht!]
\centering
\includegraphics[width=.65\textwidth]{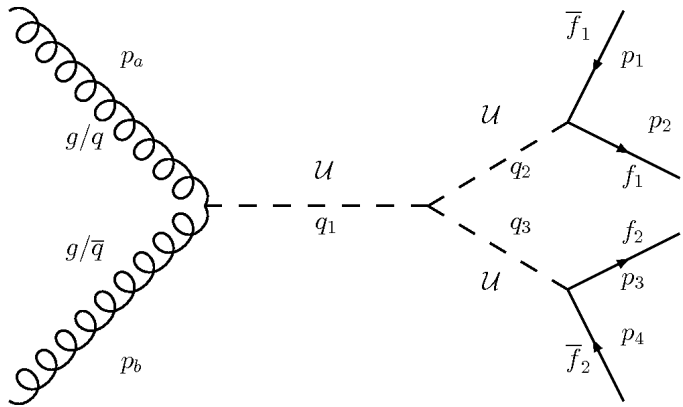}\vspace{-40.0pt}
\caption{The Feynman diagram for the subprocesses $gg/q\bar{q} \rar \U
  \rar \U \U \rar f_1 \bar{f_1}f_2 \bar{f_2}$.}
\label{fig:unpdiagram}
\end{figure}

The natural choice when dealing with these kinds of self-interactions
is to concentrate on the effects of the three-point function. First,
as discussed in Sec.~\ref{sec:tpf}, the three-point function is
completely specified up to an overall constant by conformal
invariance, while the four- and higher-point functions are not.
Second, this is the lowest order at which the resulting signal will
have a strongly suppressed background. The unparticle three-point
function can mediate processes such as
$$
pp \rar \U \rar \U\U \rar 4 \gamma, 2 \gamma 2\ell, 4\ell
$$
and many others at the LHC. The basic parton-level Feynman diagrams
are similar to the one shown in Fig.~\ref{fig:unpdiagram}. The initial
state can be either a gluon or a quark-antiquark pair, while the final
state consists of two SM particle-antiparticle pairs, i.e., two of the
pairs $gg$, $q \bar{q}$, $W^+ W^-$, $ZZ$, $\gamma \gamma$, $\ell^+
\ell^-$, $\nu \bar{\nu}$, and $HH$. In principle, one could also
couple SM particles to one of the legs in the unparticle three-point
function only, while amputating the other leg. The amputated
three-point function has the form of the ordinary one, but with the
amputated field replaced with its \emph{conformal partner} with
scaling dimension $4-d$ \cite{Mitra:2009zm}. This would lead to
processes like
$$
pp \rar \U \rar 2 \gamma \: \U, 2\ell \: \U
$$
with a phase space for two particles and one unparticle. Note that,
although the addition of unparticle lines from higher-point functions
leads to no suppression, coupling of these to SM particles does lead
to suppression simply because the particle-unparticle coupling is
assumed to be small.

In producing the results of this work, the program CompHEP
\cite{Boos:2004kh,Pukhov:1999gg} has been used. It includes the Monte
Carlo integration algorithm VEGAS \cite{Lepage:1977sw}, which is used
to integrate over phase space, and the CTEQ6L1 parton distribution
functions (PDFs) \cite{Pumplin:2002vw}. No high-energy physics program that we are aware of has the ability to incorporate such a complicated momentum dependence as the one entering through the unparticle three-point function. Therefore, in order to use the user-friendly environment of CompHEP when performing phase space and PDF integrations, we have altered the C code to be able to enter the values of the three-point function. This makes it difficult to perform more complete simulations including the interference with SM diagrams and diagrams with one or two unparticle propagators. However, including the interference with SM diagrams will not be necessary at the Tevatron, since they are negligible (for the $4 \gamma$ channel) or can be made so by imposing appropriate cuts (for the $4\ell$ channel). This also means that the backgrounds will be unimportant at the LHC since the SM-unparticle couplings become more relevant at higher energies. If $\Lambda_4$ is larger than a few TeV or $d$ is not very close to unity, inclusion of the diagrams with one or two unparticle propagators will also not be necessary. The reason is that a diagram with a single unparticle propagator contains two SM-unparticle couplings, and thus are suppressed by $1/\Lambda_4^{4d}$ in the cross section. In the same manner, two unparticle propagators means a $1/\Lambda_4^{8d}$ suppression. Of course, the contribution from the unparticle three-point function will be suppressed by $1/\Lambda_4^{6d}$, but this can be counteracted by potentially large values of $C_d$.

For leptons and photons in the final state, there are six different
squared amplitudes, corresponding to the different subprocesses \be
gg, q\bar{q} \rar \gamma \gamma \gamma \gamma, \gamma \gamma \ell
\bar{ \ell}, \ell_1 \bar{\ell}_1 \ell_2 \bar{\ell}_2. \ee When
calculating observables, the convention is to set $c_g=c_\gamma = 1$
and $e^2 (c_4^f)^2 = 2 \pi$ \cite{Bander:2007nd,Eichten:1983hw}. In
addition, following Ref.~\cite{Feng:2008ae}, $\Lambda_4 = 1 \: \tev$
is chosen. This means that the only free parameters we have left are
$d$ and $C_d$. 

It is worth emphasizing that there is no theoretical knowledge on what the values of $C_d$ are in typical models of the unparticle sector. Also, there are no model independent bounds on $C_d$ coming from more general considerations, although the authors of Ref.~\cite{Rychkov:2009ij} believe their method could be used to derive such bounds. As will be shown, generation of the relevant four-body final states requires $C_d \gg 1$. Because of the lack of knowledge on typical values of $C_d$, one cannot say if this makes it impossible to probe some models of the hidden sector through their self-interactions. In addition, if the relevant final states were to be observed at the LHC, this could not be used to extract any further information about the nature of hidden sector except the scaling dimension itself.

It is easy to observe that all cross sections will be
proportional to $C_d^2/\Lambda_4^{6d}$ and that the values of $C_d$
and $\Lambda_4$ have no effect other than an overall factor in the
cross sections. For numerical calculations, the value $C_d = 1$ is
used and can then be given the appropriate value to yield the actual
cross sections. Thus, the value of $\Lambda_4$ chosen has no effect on
the total cross sections in the above processes if a change in
$\Lambda_4$ can be accompanied by a change in the unparticle
self-coupling constant $C_d$. However, this also means that an overall
change of scale of the dimensionless couplings between the unparticle
sector and the SM can be compensated by a change in $C_d$. For the sensitivity
reach of the LHC on the unparticle couplings to photons and leptons,
see e.g. Ref.~\cite{Sahin:2009gq}. 
Because of the unitarity bound $d \geq 1$ and the more practical bound $d < 2$,
only this range on the scaling dimension is considered. More
specifically, the values $d = 1.1, 1.5, $ and $1.9$ are used as
representative values, hopefully giving a picture of the differences
caused by the different scaling dimensions in the chosen range.
\vspace{-10.0pt}

\subsection{Four-Photon Final States}

The four-photon final state mediated by unparticles has previously
been studied in Ref.~\cite{Feng:2008ae}. We include a similar analysis
for completeness and also give distributions of invariant masses,
which can be used to identify unparticles as the source as well as
identify the scaling dimension. Using the Feynman rules given in
Sec.~\ref{sec:SMcouplings}, one finds that the squared amplitudes for
the $gg \rar 4 \gamma$ and $q\bar{q} \rar 4 \gamma$ subprocesses are
\begin{align}
\bar{\lvert \mathcal{M}_{gg \rar 4\gamma} \rvert^2} &= 2^{10} \frac{c_g^2 c_\gamma^4}{\Lambda_4^{6d}} (p_a \cdot p_b)^2(p_1 \cdot p_2)^2 (p_3 \cdot p_4)^2 \lvert \Gamma_3(q_2,q_3;d) \rvert^2 , \\
 \bar{\lvert \mathcal{M}_{q\bar{q} \rar 4\gamma} \rvert^2} &= \frac{2^{9}}{3} \frac{c_\gamma^4 e^2 (c_4^f)^2}{\Lambda_4^{6d}} v^2 (p_a \cdot p_b) (p_1 \cdot p_2)^2 (p_3 \cdot p_4)^2 \lvert \Gamma_3(q_2,q_3;d) \rvert^2,
\end{align} 
respectively. These expressions do not include the factor of 1/4!
needed to compensate for the overcounting caused by the integration
over phase space for four identical particles. By dimensional
analysis, the total partonic cross sections are given by
\begin{align}
\hat{\sigma}(\hat{s})_{gg \rar 4 \gamma} &= f_d^g C_d^2 \left(\frac{\hat{s}}{\Lambda^2_4}\right)^{3d} \frac{1}{\hat{s}/ \gev^2 } \: \fb, \\
\hat{\sigma}(\hat{s})_{q\bar{q} \rar 4 \gamma} &= f_d^q C_d^2 \left(\frac{\hat{s}}{\Lambda^2_4}\right)^{3d} \frac{v^2}{\hat{s}} \frac{1}{\hat{s}/\gev^2 } \: \fb,
\end{align}
where $f_d^g$ and $ f_d^q$ are dimensionless numbers, which can be
extracted from the Monte Carlo integration and are listed in
Table~\ref{table:cstable} for different values of $d$.

The final state photons are required to satisfy the
conditions\footnote{All photons have transverse momenta well above $15
  \; \gev$. Thus, imposing slightly harder $p_T$-cuts will have no
  effect on the signal.}
\begin{equation}
\label{cuts}
p_T^{i} > 15 \; \gev, \;\;\; \lvert \eta^i \rvert < 2.5
\end{equation} 
with $p_T^i$ and $\eta^i$ being the transverse momentum and the
rapidity of the $i$-th photon, respectively, and $i = 1,2,3,4$. The
resulting reference cross sections $\sigma_{gg/q \bar{q} \to 4
\gamma}^{\text{LHC,ref}}$ (i.e., using $C_d = 1$) at the LHC with
$\sqrt{s} = 14 \; \tev$ are presented in Table~\ref{table:cstable}.
Note that our cross sections differ slightly from the results
presented in Ref.~\cite{Feng:2008ae}, which is most likely due to
different parton distribution functions. As argued in
Sec.~\ref{sec:tevatron}, Tevatron measurements give constraints on
$C_d$. Multiplying the sum of the reference cross sections by this
number squared, yields upper bounds on the LHC cross sections,
$\sigma_{4 \gamma}^{\text{LHC,max}}$, which can also be found in
Table~\ref{table:cstable}. The allowed cross sections are observed to
depend strongly on $d$, from only $6.0$~fb for $d=1.1$ to $1.2 \cdot
10^{6}$~fb for $d=1.9$.

\begin{figure}[t!]
\centering
\includegraphics[width=.60\textwidth]{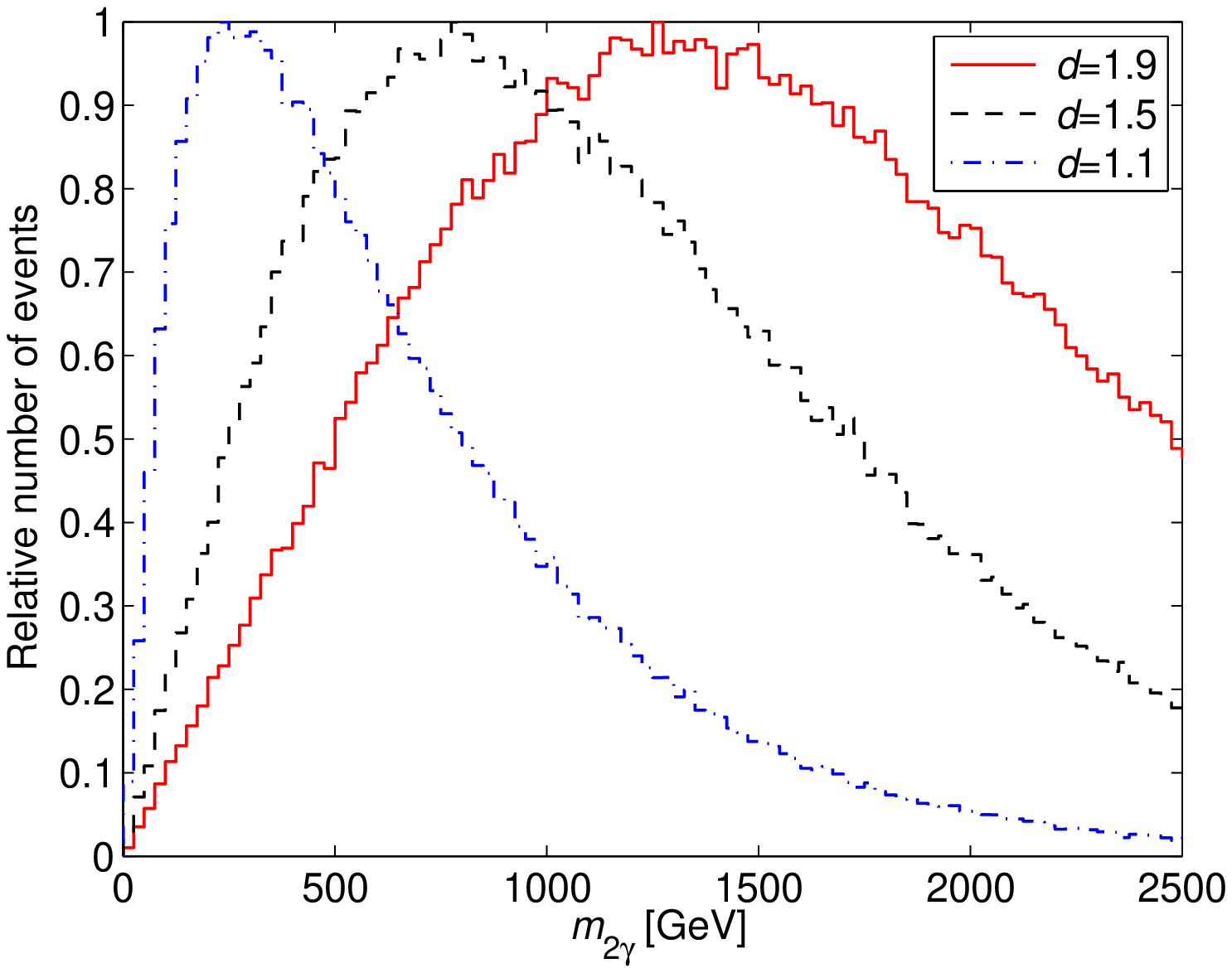} 
\caption{The distributions of photon two-body invariant masses in
  unparticle-mediated $4\gamma$ events at the LHC for three different
  values of the scaling dimension $d = 1.9$ (red solid curve), $d =
  1.5$ (black dashed curve), and $d = 1.1$ (blue dashed-dotted
  curve). The bin width used is 25~GeV. Note that each event
  contributes six counts.}
\label{fig:mij_4gamma}
\includegraphics[width=.60\textwidth]{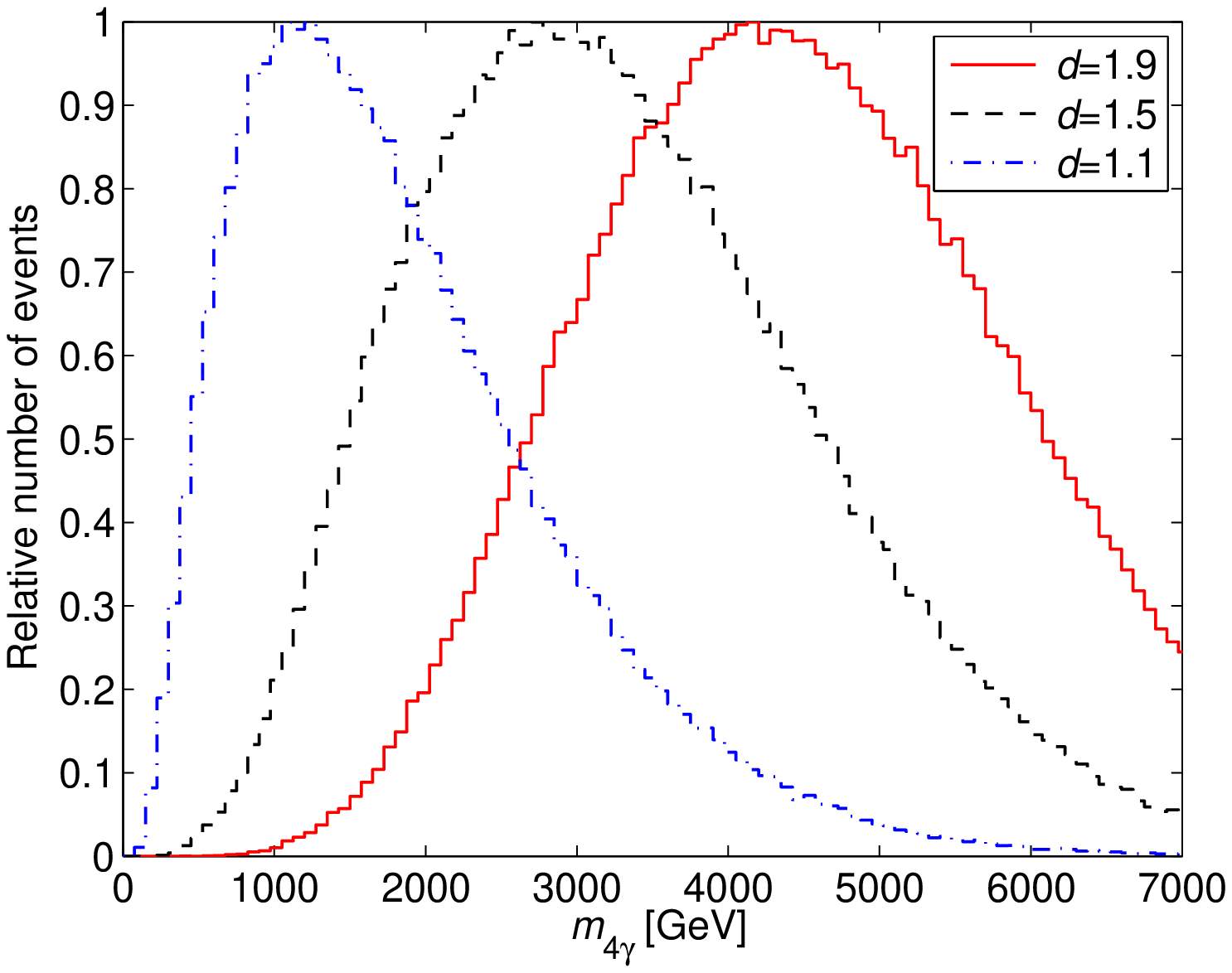} 
\caption{The distributions of photon four-body invariant masses in
  unparticle-mediated $4\gamma$ events at the LHC for three different
  values of the scaling dimension $d = 1.9$ (red solid curve), $d =
  1.5$ (black dashed curve), and $d = 1.1$ (blue dashed-dotted
  curve). The bin width used is 75~GeV.}
\label{fig:mijkl_4gamma}
\end{figure}

We have also generated a large number of events, enabling the
calculation of different distributions. In Fig.~\ref{fig:mij_4gamma},
the distribution of the two-body invariant photon masses
$m_{ij}=\sqrt{(p_i + p_j)^2}$ are shown. Since there are six different
ways of combining four identical particles, each event will contribute
six counts. In Fig.~\ref{fig:mijkl_4gamma}, the distribution of the
full four-body invariant mass $m_{4 \gamma} =
\sqrt{(p_1+p_2+p_3+p_4)^2}$ is shown. These spectra are presented for
the $gg$ initial state only, which dominate at the LHC. However, after
having generated spectra for a sample of $q \bar{q}$ initial states
and different values of $d$, one finds almost perfect agreement with
the spectra for the $gg$ initial state. Thus, if for some reason the
unparticle field does not couple to gluons at all, the resulting
spectra would be virtually identical although the total cross sections
would be different. What is evident is, first of all, the absence of
any resonance, owing to the scale-invariant nature of the unparticle
sector. Second, these distributions depend very strongly on the
unparticle scaling dimension. This is a general property of all the
spectra presented in this work; they all become very much softer when
$d$ decreases from $1.9$ to $1.1$.

The SM background consists mainly of direct diphoton production with two photons attached to the quark lines. With the same cuts as above, we find the SM background to be approximately 0.1 fb. Obviously, if the unparticle cross sections are anywhere near the values allowed by Tevatron data, there is no need to consider the interference effects with the SM. Thus, if there was to be
an excess of $4\gamma$ events at the LHC, the given distributions could
possibly be used to identify unparticles as the source, as well as the
correct value of the scaling dimension $d$. Since there is no
distinction between the couplings to photons and gluons, the
distributions of gluons in the $4g$ and $2g 2\gamma$ channels would be
identical. Although the total cross section
will be larger in this case due to the multiple colors of the gluons,
the large QCD background will make the observation in this channel
more difficult.

\begin{table}[!htb]
\begin{center}
\begin{tabular}{|c|c|c|c|}
\hline
& $d=1.1$ & $d=1.5$ & $d=1.9$ \\
\hline
\hline
$f^g_d$ & 3.0 & 0.17 & 0.021 \\
\hline
$f^q_d$ & 6.3 & 0.35 & 0.044 \\
\hline
$h^g_d$ & 790 & 16 & 1.6 \\
\hline
$h^q_d$ & $1.7 \cdot 10^{3}$ & 34 & 3.3 \\
\hline
$j^g_d$ & $1.3 \cdot 10^{4}$ & 160 & 14 \\
\hline
$j^q_d$ & $2.8 \cdot 10^{4}$ & 330 & 29 \\
\hline
\hline
Combined $4\gamma$ and $4 \ell$ bound on $C_d$ &
$450$ & $1.8 \cdot 10^4$ & $1.9 \cdot 10^5$ \\
\hline
\hline
$\sigma_{g g \to 4 \gamma}^{\text{LHC,ref}} $ [fb] 
& $2.9 \cdot 10^{-5} $ & $1.1 \cdot 10^{-5} $ & $3.2 \cdot 10^{-5}$ \\
\hline
$\sigma_{q \bar{q} \to 4 \gamma}^{\text{LHC,ref}} $ [fb]
& $1.0 \cdot 10^{-6}$ & $3.8 \cdot 10^{-7}$ & $1.1 \cdot 10^{-6}$ \\
\hline
$\sigma_{4 \gamma}^{\text{LHC,max}} $ [fb]
& $6.0$ & $3.7 \cdot 10^{3}$ & $1.2 \cdot 10^{6}$ \\
\hline
\hline
$\sigma_{g g \to 2 \gamma 2 \ell}^{\text{LHC,ref}} $ [fb]
& $3.8 \cdot 10^{-4}$ & $1.1 \cdot 10^{-5}$ & $1.0 \cdot 10^{-5}$ \\
\hline
$\sigma_{q \bar{q} \to 2 \gamma 2 \ell}^{\text{LHC,ref}} $ [fb]
& $2.0\cdot 10^{-5}$ & $4.0 \cdot 10^{-7}$ & $3.7 \cdot 10^{-7}$ \\
\hline
$\sigma_{2 \gamma 2 \ell}^{\text{LHC,max}} $ [fb]
& $81$ & $3.7 \cdot 10^{3}$ & $3.7 \cdot 10^{5}$ \\
\hline
\hline
$\sigma_{g g \to 2 e 2 \mu}^{\text{LHC,ref}} $ [fb]
& $ 2.6 \cdot 10^{-3} $ & $4.6 \cdot 10^{-6}$ & $7.7 \cdot 10^{-7}$ \\
\hline
$\sigma_{q \bar{q} \to 2 e 2 \mu}^{\text{LHC,ref}} $ [fb]
& $2.7 \cdot 10^{-4}$ & $2.2 \cdot 10^{-7}$ & $2.6 \cdot 10^{-8}$ \\
\hline
$\sigma_{2 e 2 \mu}^{\text{LHC,max}} $ [fb]
& $580$ & $1.6 \cdot 10^{3}$ & $2.9 \cdot 10^{4}$ \\
\hline
\end{tabular}
\end{center}
\caption{The dimensionless constants $f^{g,q}_d$, $h^{g,q}_d$, and
  $j^{g,q}_d $, appearing in the partonic cross sections, as well as
  the unparticle $4\gamma$, $2\gamma 2 \ell$, and $2 e 2 \mu$
  reference cross sections at the LHC for $d=1.1, 1.5$, and $1.9$.
  The reference cross sections are calculated with $C_d = 1$, while
  the actual cross sections are proportional to $C_d^2$, which is
  bounded by Tevatron data (see Sec.~\ref{sec:tevatron}), giving upper
  bounds on total cross sections for all final states at the LHC.}
\label{table:cstable}
\end{table}

\subsection{Two-Photon and Two-Lepton Final States}
\label{sec:2gamma2l}

Simply replacing two of the photons with one charged lepton-antilepton
pair, yields the subprocesses $gg \rar 2 \gamma 2\ell$ and $q\bar{q}
\rar 2 \gamma 2\ell$ with squared amplitudes
\begin{align} \label{eq:M2:2gamma2l}
\bar{\lvert \mathcal{M}_{gg \rar 2 \gamma 2\ell} \rvert^2} &=64 \frac{c_\gamma^2 c_g^2 e^2 (c_4^f)^2}{\Lambda_4^{6d}} v^2 (p_a \cdot p_b)^2 (p_1 \cdot p_2)^2 (p_3 \cdot p_4) \lvert \Gamma_3(q_2,q_3;d) \rvert^2 , \\
\bar{\lvert \mathcal{M}_{q\bar{q} \rar 2 \gamma 2\ell} \rvert^2} &=\frac{32}{3} \frac{c_\gamma^2 e^4 (c_4^f)^4}{\Lambda_4^{6d}} v^4 (p_a \cdot p_b) (p_1 \cdot p_2)^2 (p_3 \cdot p_4) \lvert \Gamma_3(q_2,q_3;d) \rvert^2,
\end{align} 
respectively. These expressions should be multiplied with 1/2! to
account for the identical photons in the final state. Just as in the
four-photon case, the partonic cross sections can be written as
\begin{align}
\hat{\sigma}(\hat{s})_{gg \rar 2\gamma 2 \ell} &= h_d^g C_d^2 \left(\frac{\hat{s}}{\Lambda_4^2}\right)^{3d} \frac{v^2}{\hat{s}} \frac{1}{\hat{s}/\gev^2 } \: \fb, \\
\hat{\sigma}(\hat{s})_{q\bar{q} \rar 2\gamma 2 \ell} &= h_d^q C_d^2 \left(\frac{\hat{s}}{\Lambda_4^2}\right)^{3d} \left( \frac{v^2}{\hat{s}}\right)^2 \frac{1}{\hat{s}/\gev^2 } \: \fb,
\end{align}
where the proportionality factors are again listed in
Table~\ref{table:cstable}. The same procedure as in the four-photon
case also yields the reference and maximum cross sections at the LHC,
where the same single-particle cuts have been imposed. The allowed
cross sections are of the same order of magnitude as the $4
\gamma$ channel, although the difference between different values of
the scaling dimension is now smaller. Still, there is the possibility
to have very large cross sections for large $d$.

\begin{figure}[t!]
\centering
\includegraphics[width=.60\textwidth]{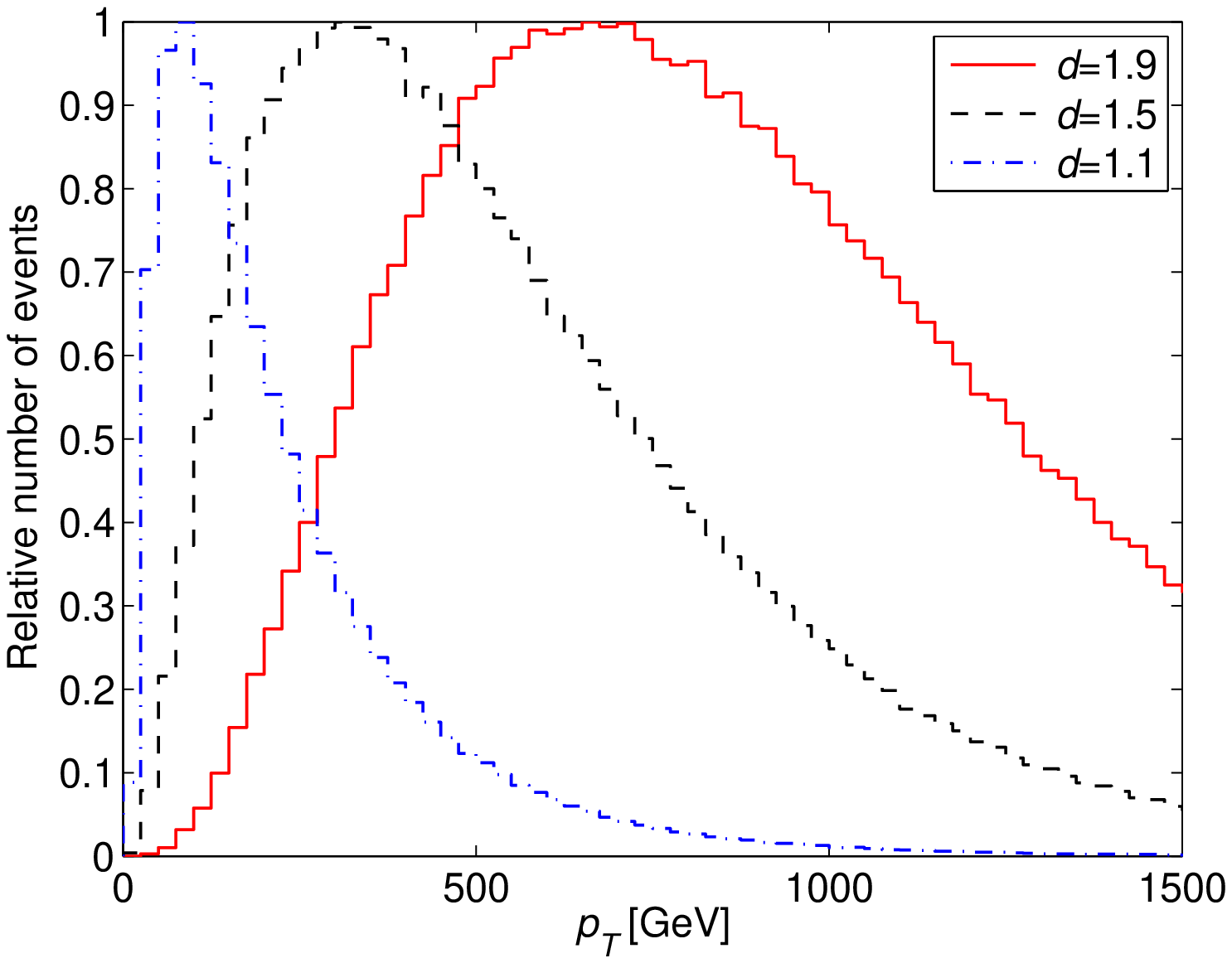}
\caption{The $p_T$ distributions of the highest-$p_T$ photon in
  unparticle-mediated 2$\gamma 2\ell$ events at the LHC for three
  different values of the scaling dimension $d = 1.9$ (red solid
  curve), $d = 1.5$ (black dashed curve), and $d = 1.1$ (blue
  dashed-dotted curve). The bin width used is 25~GeV.}
\label{fig:ptmaxgamma_2l2gamma}
\includegraphics[width=.60\textwidth]{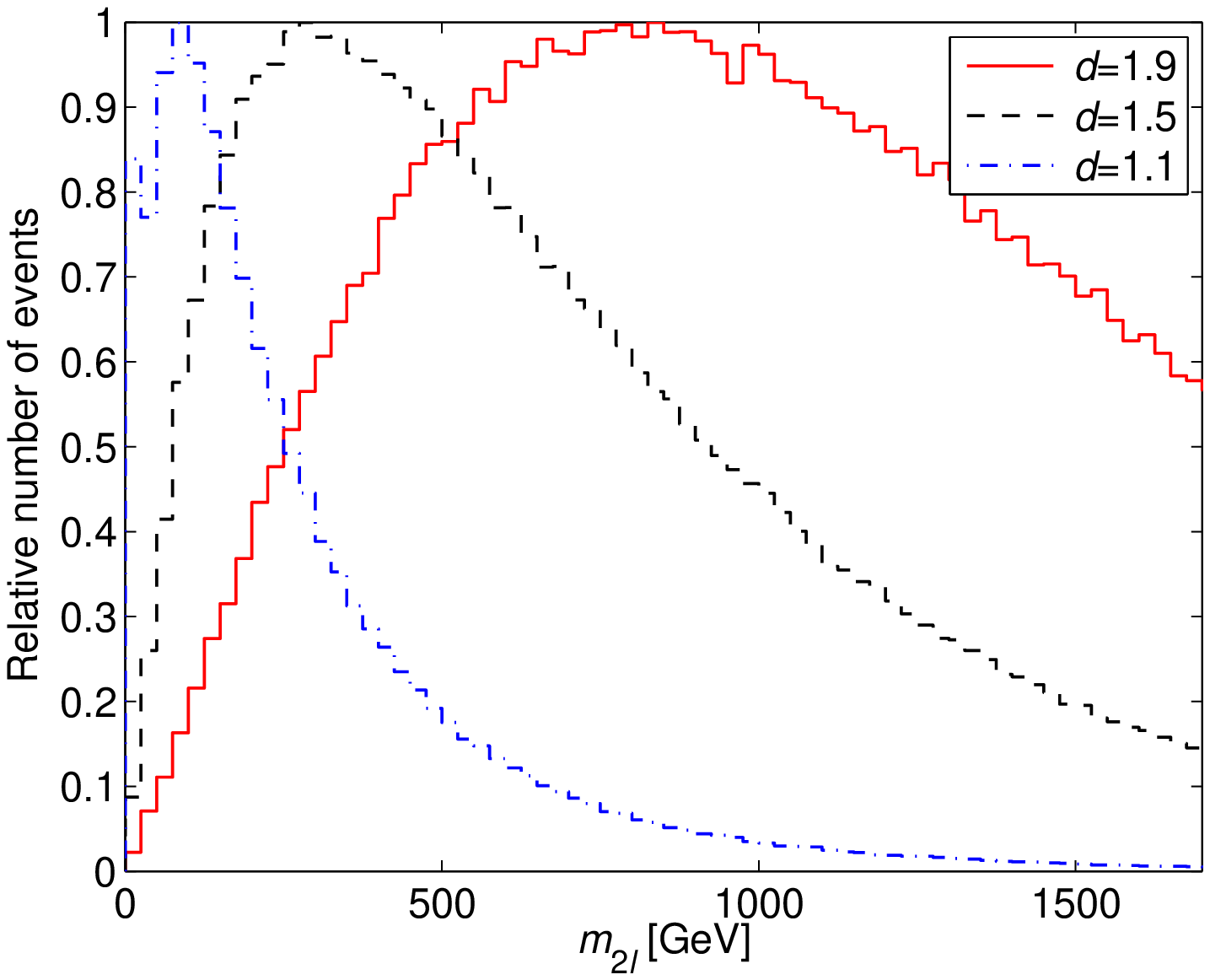}
\caption{The distributions of the charged lepton invariant masses in
  unparticle-mediated 2$\gamma 2\ell$ events at the LHC for three
  different values of the scaling dimension $d = 1.9$ (red solid
  curve), $d = 1.5$ (black dashed curve), and $d = 1.1$ (blue
  dashed-dotted curve). The bin width used is 25~GeV.}
\label{fig:mll_2l2gamma}
\end{figure}

The $p_T$ distributions of the photon with the highest transverse
momentum and the distribution of the charged lepton invariant masses
are shown in Figs.~\ref{fig:ptmaxgamma_2l2gamma} and
\ref{fig:mll_2l2gamma}, respectively. Again, one observes the
dramatic softening of the distributions as the scaling dimension
decreases and just as in the four-photon case, almost perfect
agreement between the distributions from the $gg$ and $q \bar{q}$
initial states are found.

Using CompHEP, the SM cross section is calculated to be of the order of 30 fb per lepton flavor. There will be resonances at charged lepton invariant masses $0$ and $M_Z \simeq 90 \; \gev$. Thus, we demand that the charged lepton invariant masses exceed 110 GeV, which yields a reduction of the cross section to approximately $1 \: \fb$. As can be observed from Fig.~\ref{fig:mll_2l2gamma}, this cut will have a very small effect on the signal, especially for $d$ not very close to unity. Accordingly, the SM backgrounds will be negligible for $C_d$ anywhere close to its allowed values. In analogy with the four-photon case, since the unparticle field is
assumed to couple to quarks and gluons, one should also find $q
\bar{q} g g$, $q \bar{q} \gamma \gamma$, and $\ell \bar{\ell} gg$
final states with identical spectra, but larger cross sections.

\subsection{Four-Charged Lepton Final States}
\label{sec:4l}

For four charged leptons in the final state, the squared amplitudes
are given by
\begin{align}
\label{eq:M2:4l}
\bar{\lvert \mathcal{M}_{gg \rar 4 \ell} \rvert^2} &= 4 \frac{c_g^2
e^4 (c_4^f)^4 }{\Lambda_4^{6d}} v^4 (p_a \cdot p_b)^2 (p_1 \cdot
p_2) (p_3 \cdot p_4) \lvert \Gamma_3(q_2,q_3;d) \rvert^2 ,
\\ \bar{\lvert \mathcal{M}_{q\bar{q} \rar 4 \ell} \rvert^2} &=
\frac{2}{3} \frac{e^6 (c_4^f)^6 }{\Lambda_4^{6d}} v^6 (p_a \cdot p_b)
(p_1 \cdot p_2) (p_3 \cdot p_4) \lvert \Gamma_3(q_2,q_3;d)
\rvert^2,
\end{align}
respectively. The relevant final states are $4e$, $4 \mu$ and $2 e 2
\mu$. The cross sections can again be found in
Table~\ref{table:cstable} and are given for the $2 e 2 \mu$ final
state. Similar to the previous final states, the partonic cross
sections are
\begin{align}
\hat{\sigma}(\hat{s})_{gg \rar 4 \ell} &= j_d^g C_d^2 \left(\frac{\hat{s}}{\Lambda_4^2}\right)^{3d} \left( \frac{v^2}{\hat{s}} \right)^2 \frac{1}{\hat{s}/\gev^2 } \: \fb, \\
\hat{\sigma}(\hat{s})_{q\bar{q} \rar 4 \ell } &= j_d^q C_d^2 \left(\frac{\hat{s}}{\Lambda_4^2}\right)^{3d} \left( \frac{v^2}{\hat{s}} \right)^3 \frac{1}{\hat{s}/\gev^2 } \: \fb .
\end{align}

\begin{figure}[t!]
\centering	 
\includegraphics[width=.60\textwidth]{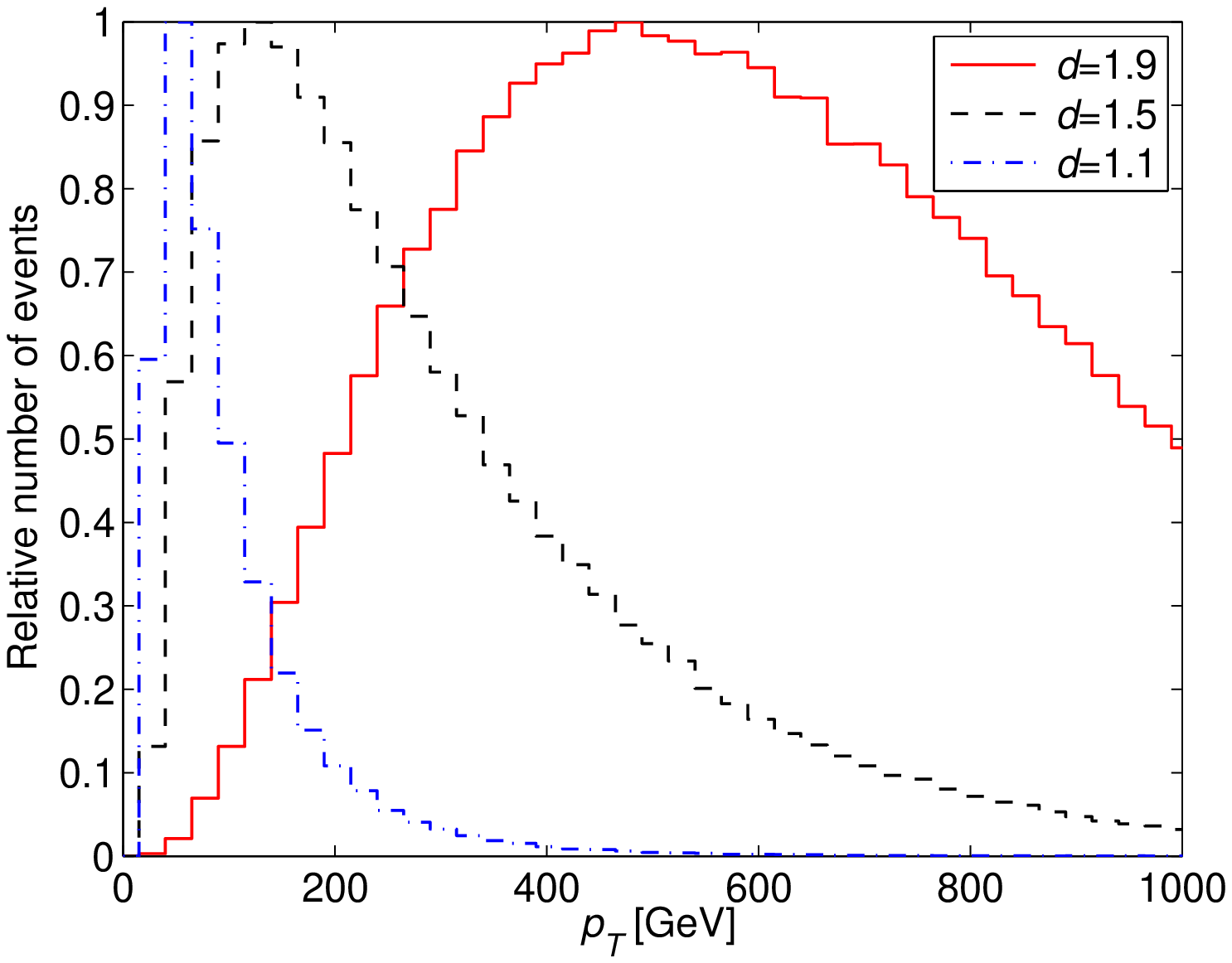} 
\caption{The $p_T$-distributions of the highest-$p_T$ lepton in
  unparticle-mediated $4\ell$ events at the LHC for three different
  values of the scaling dimension $d = 1.9$ (red solid curve), $d =
  1.5$ (black dashed curve), and $d = 1.1$ (blue dashed-dotted
  curve). The bin width used is 25~GeV.}
\label{fig:ptmax_4l}
\includegraphics[width=.60\textwidth]{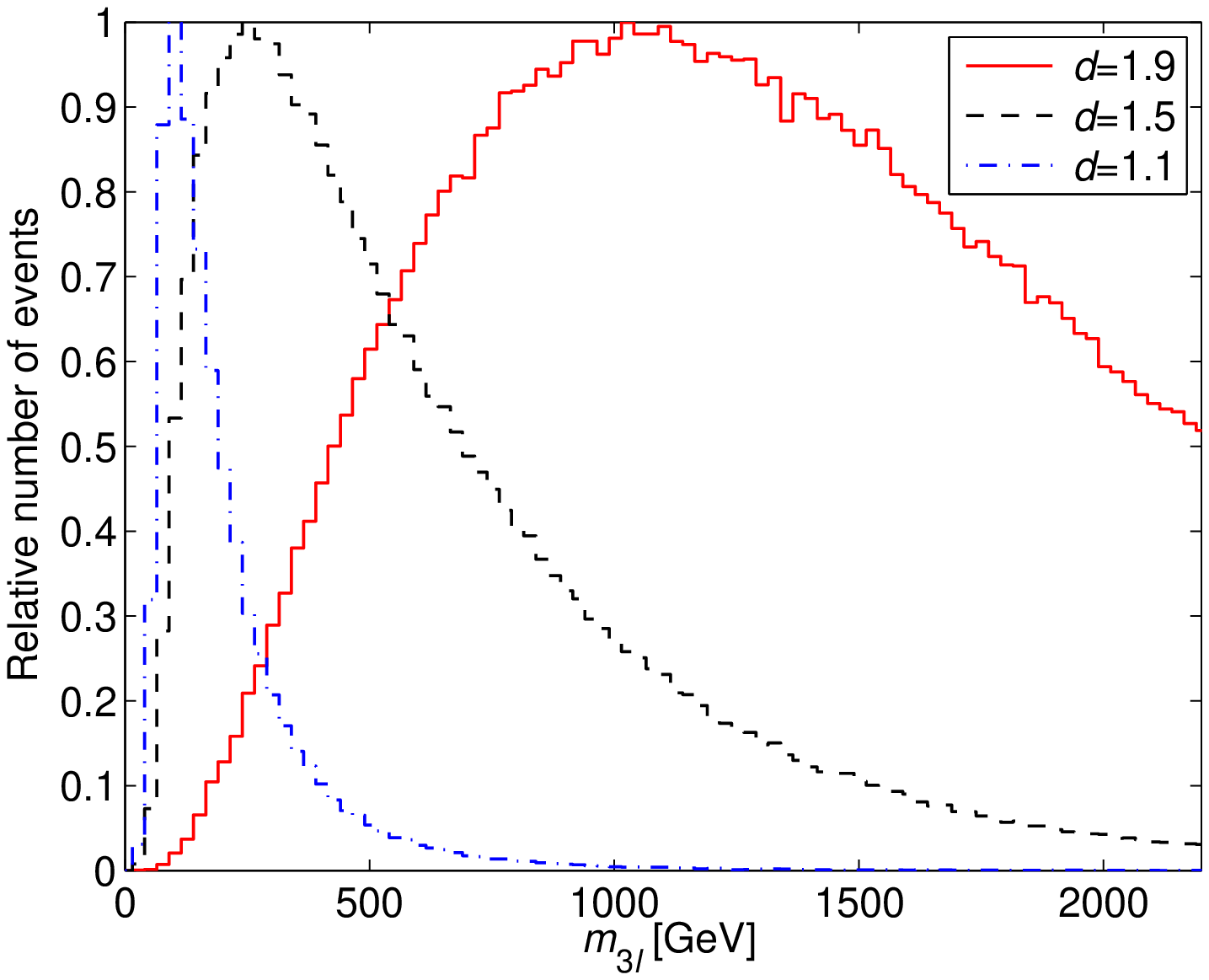}
\caption{The distributions of the three-body invariant lepton masses
  in unparticle-mediated $4\ell$ events at the LHC for three different
  values of the scaling dimension $d = 1.9$ (red solid curve), $d =
  1.5$ (black dashed curve), and $d = 1.1$ (blue dashed-dotted
  curve). The bin width used is 25~GeV. Note that each event
  contributes two counts.}
\label{fig:mlll_4l}
\end{figure}

In Fig.~\ref{fig:ptmax_4l}, the $p_T$ distributions of the
highest-$p_T$ lepton are shown, while Fig.~\ref{fig:mlll_4l} depicts
the distribution of the invariant masses of three leptons of which two
have the same charge. Since there are two ways of combining three
leptons, each event will contribute two counts. Again, there is a
strong dependence on $d$.

The four-charged lepton channel is a very clean and important channel
when it comes to the discovery potential of the Higgs boson ($H
\rightarrow ZZ \rightarrow 4 \ell$) at the LHC. Thus, one can compare
the cross sections originating from the unparticle self-interactions
to the already existing calculated backgrounds in connection with the
Higgs boson \cite{ATLAS2008-013,ATLAS2008-045}. With the cuts $p_T > 5
\; \gev$ and $|\eta| < 2.5$, the irreducible background is about
$70~\fb$, coming mainly from two $Z$-bosons decaying into four charged
leptons. Furthermore, with cuts on the invariant masses of the charged leptons, it should be easy to reduce this part of the background to a few fb. The reducible backgrounds stem from the $t\bar{t} \rar W^+ b
W^- \bar{b} \rar 4\ell + X $ and $Z b \bar{b} \rar 4\ell + X $
processes with a cross section of $7.0 \cdot 10^3$~fb. Since the
events coming from these processes contain $b$ jets in the final
state, they can be reduced below the irreducible levels, using cuts on
the track isolation and impact parameters without significant
reduction of the signal. Thus, the possible cross sections from
unparticles summed over all channels, $10^3 \; \fb - 10^5 \; \fb $,
are potentially huge compared to the backgrounds.

\subsection{Tevatron Bounds}
\label{sec:tevatron}

In order to give bounds on the cross sections of processes mediated by
unparticle self-interactions, one needs bounds on the constant $C_d$
appearing in the three-point function in \eref{eq:3point}, which
determines the strength of the self-interaction. However, there is no
known way to bound $C_d$ theoretically, and thus, one needs to turn to
experiments. These bounds could come from many different
experiments. Since unparticle effects are most easily probed at
high-energy colliders and there are no LHC data available yet, it is
natural to turn to the existing Tevatron data. It is important to
remember that these bounds, and hence, the bounds on the LHC cross
sections, depend on the values given to the dimensionless coupling
constants between the SM and unparticle fields. For example, if the
unparticle field does not couple to gluons at all, the allowed cross
section at the LHC will be smaller, since the $gg$ initial state is
dominating at the LHC, while the $q \bar{q}$ initial state is
dominating at the Tevatron.

\subsubsection{The 4$\gamma$ channel}

In this section, we repeat the analysis performed in
Ref.~\cite{Feng:2008ae}, but find less stringent constraints on $C_d$.
This is because we find smaller cross sections\footnote{However, the
  cross sections are almost identical to the ones in
  Ref.~\cite{Feng:2008ae} when we remove all our cuts.} and we take
into account the single photon detection efficiency.

The SM background at the Tevatron is calculated to be approximately 0.01 fb.
Using $0.83 \; \fb^{-1}$ of data, the D0 Collaboration found no
candidate events \cite{D05067}. Thus,
the 95~\% C.L.~bound equal to three on the mean of a Poisson process
\cite{Feldman:1997qc} becomes the bound on $C_d$
$$C_d^2 < \frac{3}{0.83 \: \fb^{-1} \sigma_{4 \gamma}^{\text{Tev,ref}} \epsilon_\gamma^4 },
$$
where $\sigma_{4 \gamma}^{\text{Tev,ref}}$ is the total reference
cross section at the Tevatron and $\epsilon_\gamma = 0.9$ is taken
as the single photon detection efficiency. The values of $\sigma_{4
\gamma}^{\text{Tev,ref}}$ and the bounds on $C_d$ can be found in
Table~\ref{table:4gamma}. These bounds are approximately 4 times
less restrictive in terms of cross sections than the bounds found in
Ref.~\cite{Feng:2008ae}. However, this does not matter, since the
bounds from the $4 \ell$ channel calculated in
Sec.~\ref{sec:4lchannel} will be even more restrictive.

\begin{table}[!htb]
\begin{center}
\begin{tabular}{|c|c|c|c|}
\hline
& $d=1.1$ & $d=1.5$ & $d=1.9$ \\
\hline
\hline
$\sigma_{4 \gamma}^{\text{Tev,ref}}$ [fb]
& $8.5 \cdot 10^{-9}$ & $1.0 \cdot 10^{-10}$ & $ 6.1 \cdot 10^{-12}$ \\
\hline
\hline
Bounds on $C_d$ & $2.6 \cdot 10^{4}$ & $2.3 \cdot 10^{5}$ & $9.5 \cdot 10^{5}$ \\
\hline
\end{tabular}
\end{center}
\caption{The Tevatron reference cross sections in the $4 \gamma$
  channel and the resulting upper bounds on $C_d$.}
\label{table:4gamma}
\end{table}

\subsubsection{The $4\ell$ channel}
\label{sec:4lchannel}

The D0 Collaboration has also searched for four-charged lepton final
states \cite{:2007hm,Jarvis:2007zz,D05345} with an approximate
integrated luminosity of $1 \: \mathrm{fb^{-1}}$ in all
channels. Since electrons and muons are detected in different ways in
the detector, one needs to separate their analyses.

For the $4 \mu $ channel, the relevant geometric cut is $|\eta| < 2 $
and the kinematical ones are $p_T > 15 \; \gev$ and $\cos{\alpha_i} <
0.96$, where the $\alpha_i$'s are the angles between all six muon
pairs. Also, the two invariant masses of the muon pairs in at least
one of three possible pairings must both be larger than 30 GeV. In
addition, there are losses in acceptances due to the requirement on
having a track match, the quality of the signal of the detected muons,
and the efficiencies of the triggers. We implement the cuts on $\eta$,
$p_T$, $\cos{\alpha_i}$ and also the more conservative demand that the
invariant masses of $M_{12}=\sqrt{(p_1+p_2)^2}$ and
$M_{34}=\sqrt{(p_3+p_4)^2}$ are both larger than 30 GeV. To eliminate
the SM background, which is dominated by the intermediate states with
two $Z$ bosons or photons, the additional cut $M_{12}, M_{34} \notin
\left[ M_Z - 4 \Gamma_Z , M_Z + 4 \Gamma_Z \right] = \left[ 81.2 \:
  \gev, \: 101.2 \: \gev \right] $ is imposed. We approximate the
acceptances for the track match, detector quality cuts, and trigger
efficiencies to be the same as for the $ZZ$ intermediate state in
Refs. \cite{:2007hm,Jarvis:2007zz,D05345}, giving a total acceptance
loss of approximately $\epsilon_{4 \ell} = 0.5$.

In the $4 e $ channel, the same requirements on $p_T$ and the
invariant masses are applied. However, the rapidity cuts for electrons
are instead $|\eta| < 1.1$ or $1.5 < |\eta| < 3.2$. The spatial
separation $R = \sqrt{(\eta_i - \eta_j)^2 + (\phi_i - \phi_j)^2} $ is
required to be larger than 0.4. In the same way as for the $ 4\mu$
channel, the additional acceptance loss is approximated to be 0.5 from
the cut on $R$, the electron quality cuts, and the efficiencies of the triggers.

Finally, for the $2 e 2 \mu $ channel, the same single-particle cuts
as in the $4 e$ and $4 \mu$ channels are applied. Here $\cos{\alpha} <
0.96$ for the two muons and $R > 0.2$ for all the final state
particles. The same procedure for the $ 4 e$ channel yields an
additional acceptance loss of 0.5.

The SM background is now approximately zero and no events with our
requirements on the invariant masses were found. Thus, we obtain the
95~\% C.L.~bound on $C_d$
$$
C_d^2 < \frac{3}{1 \: \fb^{-1} \epsilon_{4 \ell} \left(\sigma_{4 e}^{\text{Tev,ref}} + \sigma_{4 \mu}^{\text{Tev,ref}} + \sigma_{2 e 2 \mu}^{\text{Tev,ref}} \right) }.
$$
The cross section in each channel and the calculated upper bounds on $
C_d $ can be found in Table~\ref{table:4l}. One observes that the
bounds from the $4 \ell$ channels are about $3.3 \cdot 10^3$, 160, and
25 times more severe than the ones from the $4\gamma$ channel in terms
of cross sections for $d=1.1, 1.5 $, and 1.9, respectively. Thus, the
$4\ell$ final states produce more restrictive upper bounds on the
cross sections than the $4\gamma$ final states. In conclusion, this
means that the upper bounds on the cross sections from the $4\ell$
final states should replace the ones obtained earlier in the
literature \cite{Feng:2008ae}.
\begin{table}[!htb] 
\begin{center}
\begin{tabular}{|c|c|c|c|}
\hline
& $d=1.1$ &$d=1.5$ & $d=1.9$ \\
\hline
\hline
$\sigma_{4 \mu}^{\text{Tev,ref}}$ [fb]
& $4.6 \cdot 10^{-6}$ & $4.7 \cdot 10^{-9}$ & $4.4 \cdot 10^{-11}$ \\
\hline
$\sigma_{4 e}^{\text{Tev,ref}}$ [fb]
& $3.9 \cdot 10^{-6}$ & $1.8 \cdot 10^{-9}$ & $1.8 \cdot 10^{-11}$ \\
\hline
$\sigma_{2 e 2 \mu}^{\text{Tev,ref}}$ [fb]
& $2.1 \cdot 10^{-5}$ & $1.2 \cdot 10^{-8}$ & $1.1 \cdot 10^{-10}$ \\
\hline 
\hline
Bounds on $C_d$ & 450 & $1.8 \cdot 10^{4}$ & $1.9 \cdot 10^{5}$ \\
\hline
\end{tabular}
\end{center}
\caption{The Tevatron reference cross sections in the $4 \ell$
  channels and the calculated upper bounds on $C_d$.}
\label{table:4l}
\end{table}

\subsection{Missing Transverse Momentum}

The couplings of SM neutrinos to a scalar unparticle field are usually
taken as
\cite{Zhou:2007zq,Barranco:2009px,Montanino:2008hu,Balantekin:2007eg}
\begin{equation}
\label{eq:neucoup}
\frac{\lambda^{\alpha\beta}}{\Lambda_3^{d-1}} \bar{\nu}_\alpha
\nu_\beta \OO_{\U} ,
\end{equation}
where $\alpha, \beta = e, \mu, \tau$ are flavor indices. For flavor
conserving couplings, the matrix $\lambda = (\lambda^{\alpha\beta})$
is, of course, diagonal. This can lead to the decay of heavy neutrinos to
lighter ones \cite{Zhou:2007zq} and other interesting phenomenological
consequences. Here, however, the focus will be on processes such as
\be pp \rar 2 \gamma \nu \bar{\nu}, 2\ell \nu \bar{\nu}. \ee The
couplings in \eref{eq:neucoup} will lead to the same squared
amplitudes as given in Eqs.~(\ref{eq:M2:2gamma2l}) and
(\ref{eq:M2:4l}), up to an overall factor.

\begin{figure}[t!]
\centering
\includegraphics[width=.60\textwidth]{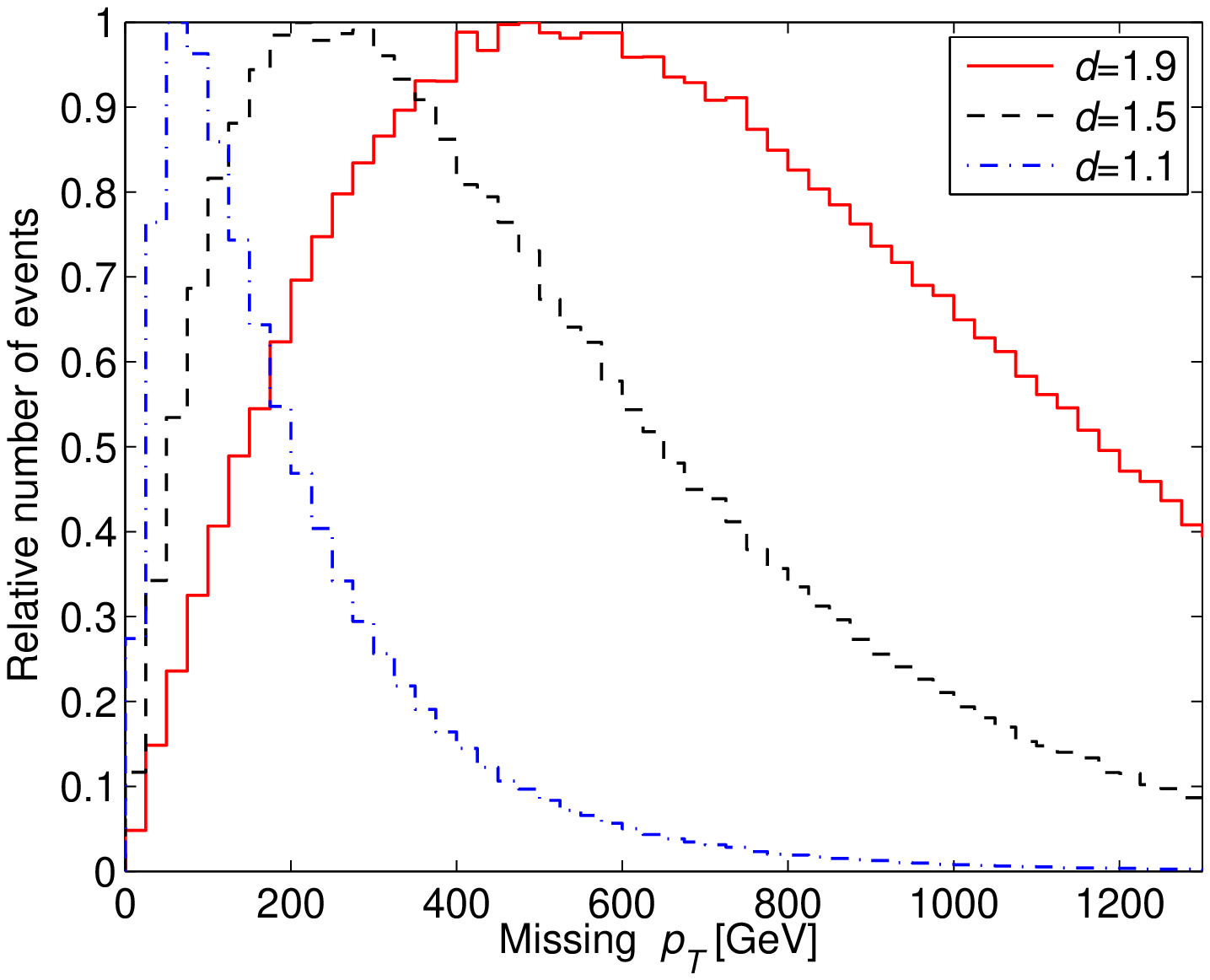}
\caption{The missing transverse momentum in unparticle-mediated $2
  \gamma \nu \bar{\nu}$ events at the LHC for three different values
  of the scaling dimension $d = 1.9$ (red solid curve), $d = 1.5$
  (black dashed curve), and $d = 1.1$ (blue dashed-dotted curve). The
  bin width used is 25~GeV.}
\label{fig:2gammapt}
\includegraphics[width=.60\textwidth]{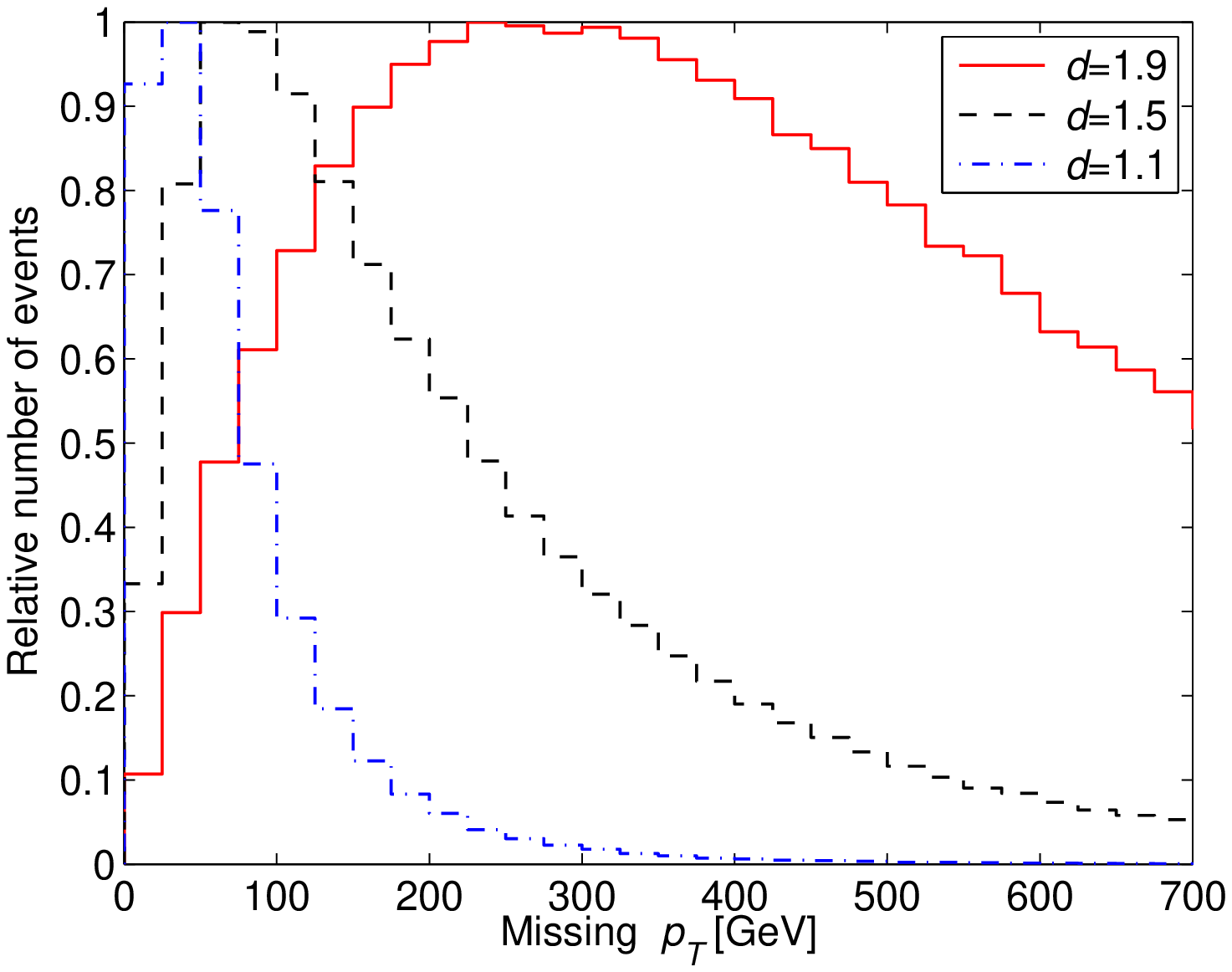}
\caption{The missing transverse momentum in unparticle-mediated $2\ell
  \nu \bar{\nu}$ events at the LHC or three different values of the
  scaling dimension $d = 1.9$ (red solid curve), $d = 1.5$ (black
  dashed curve), and $d = 1.1$ (blue dashed-dotted curve). The bin
  width used is 25~GeV. Note the different scale on the
  $\slashed{p}_T$ axis as compared to
  Fig.~\ref{fig:2gammapt}.}
\label{fig:2lpt}
\end{figure}

One cannot say anything about the total cross sections for these kinds
of processes unless one has knowledge about the couplings
$\lambda^{\alpha\beta}$. For $\lambda^{\alpha\beta}$ of the order of
unity and $\Lambda_3$ of the order of 1 TeV, the resulting cross
sections would be of the same order of magnitude as the ones
calculated for the $2\gamma 2\ell$ and $4\ell$ channels in
Secs.~\ref{sec:2gamma2l} and \ref{sec:4l}, respectively. The
distributions of missing transverse momentum, $\slashed{p}_T$, for $2
\gamma 2 \nu$ and $2\ell 2 \nu$ events at the LHC are shown in
Figs.~\ref{fig:2gammapt} and \ref{fig:2lpt}, respectively. Compared
with the events expected from SM processes
\cite{ATLAS2009-086,ATLAS2004-028}, the events mediated by unparticles
are observed to contain large $\slashed{p}_T$, especially for larger
values of $d$. In addition, since the total rates could be relatively
large, a high cut on $\slashed{p}_T$ could be used to search for
unparticle physics in the $2\gamma + \slashed{p}_T$ and $2\ell +
\slashed{p}_T$ channels as well.

\section{Summary and Conclusions}
\label{sec:summary}

Unparticle physics is a relatively new idea for an extension of the SM
of particle physics, introduced by Georgi in 2007
\cite{Georgi:2007ek}. The basic idea is that there might exist a
hidden conformal sector, which is coupled to the SM through effective
operators. Thus, unparticle physics is the study of the
phenomenological consequences of the SM interactions with such a
hidden sector.

In this work, the implications of the unparticle self-interactions,
entering through the three-point correlation function, for LHC
phenomenology have been investigated. For weak enough couplings of the unparticle sector to the SM, i.e., for a high enough unparticle scale $\Lambda_4$, unparticle physics will be impossible to observe in all previously studied processes, i.e., those including unparticle propagators and external lines. Because of the new degree of freedom entering through the self-interactions, this does not necessarily apply to the processes studied in this work, assuming large values of $C_d$ are allowed. Thus, it is not only possible that these processes could be used as a favorable way to look for unparticle physics, but also that it might be the only way.

The final states examined were
the $4 \gamma$, $2 \gamma 2\ell$, and $4\ell$ final states, the first
of which has been studied previously in the literature
\cite{Feng:2008ae}. In addition, we have used the analysis of the $4
\ell$ channel at the Tevatron to calculate constraints on the constant
appearing in the unparticle three-point correlation function, enabling
determination of the maximum allowed cross sections at the LHC. The
results take the form of upper bounds on the LHC cross sections and
distributions of transverse momenta, invariant masses, and missing
transverse momenta.

The allowed cross sections have been found to depend strongly on the
scaling dimension $d$ of the unparticle sector. In general, they are
much smaller for $d$ close to 1 and increase as $d$ increases towards
2 in which case the cross section can be as large as $10^6$~fb. For $d
= 1.9$, we obtained $10^6$~fb for the $4\gamma$ final state,
$10^5$~fb for each of the the $2\gamma 2\ell$ final states, and $10^4$~fb for each of the
$4\ell$ final states. However, for smaller values of $d$, the allowed
cross sections are much smaller due to our new constraints from the
$4\ell$ channel at the Tevatron, which are indeed much stronger and
replace the upper bounds from the $4\gamma$ channel. In addition, the
calculated distributions have been observed to depend strongly on the
scaling dimension and become much softer when it decreases towards
1. Finally, if there was to be an excess of events of the relevant
final states at the LHC (possibly enormous), unparticle physics would
be one possible description. The computed distributions could then
potentially be used to identify unparticles as the source, as well as
the correct value of the scaling dimension $d$.

\begin{acknowledgments}
We would like to thank Chad Jarvis and He Zhang for useful
discussions. This work was supported by the Royal Swedish Academy of
Sciences (KVA) [T.O.] and the Swedish Research Council
(Vetenskapsr{\aa}det), contract nos.~621-2005-3588 and 621-2008-4210
[T.O.]
\end{acknowledgments}


\end{document}